\documentclass[letterpaper]{article}

\usepackage{amsmath}
\usepackage{amsfonts}
\usepackage{mathtools}
\usepackage[justification=centering]{subcaption}
\usepackage{standalone}
\usepackage{gnuplot-lua-tikz}
\usepackage{natbib,alifeconf}

\usepackage{tikz}
\usetikzlibrary{arrows}
\newcommand\crule[3][black]{\textcolor{#1}{\rule{#2}{#3}}}
\definecolor{color1}{RGB}{27,158,119}
\definecolor{color2}{RGB}{217,95,2}
\definecolor{color3}{RGB}{117,112,179}
\definecolor{color4}{RGB}{231,41,138}
\definecolor{color5}{RGB}{102,166,30}

\title{Informational parasites in code evolution}
\author{Andr\'{e}s C. Burgos$^1$ \and Daniel Polani$^2$ \\
\mbox{}\\
$^{1,2}$Adaptive Systems Research Group, University of Hertfordshire, Hatfield, UK \\
$^1$a.c.burgos@herts.ac.uk}

\begin{document}
\maketitle

\begin{abstract}
In a previous study, we considered an information-theoretic model of code evolution. In this model, agents bet on (common) environmental
conditions using their sensory information as well as that obtained from messages of other agents, which is determined by
an interaction probability (the structure of the population). For an agent to understand another agent's messages, the former must
either know the identity of the latter, or the code producing the messages must be universally interpretable.

A universal code, however, introduces a vulnerability: a parasitic entity can take advantage of it. Here, we investigate this
problem. In our specific setting, we consider a parasite to be an agent that tries to inflict as much damage as possible in the
mutual understanding of the population (\emph{i.e.} the parasite acts as a disinformation agent). We show that, after introducing
a parasite in the population, the former adopts a code such that it captures the information about the environment that is missing
in the population. Such an agent would be of great value, but only if the rest of the population can understand its messages. However,
it is of little use here, since the parasite utilises the most common messages in the population to express different concepts.

Now we let the population respond by updating their codes such that, in this arms race, they again maximise their mutual
understanding. As a result, there is a code drift in the population where the utilisation of the messages of the parasite is avoided.
A consequence of this is that the information that the parasite possesses but which the agents lack becomes understandable and readily available.
\end{abstract}

\section{Introduction}
\label{sec:introduction}

Codes shared among entities are ubiquitous in nature, not only present in biological systems, but also, at the least, in technological ones \citep{Doyle2010}.
We define a code as a probabilistic mapping from an ``input'' random variable (\emph{e.g.} environmental variable) to a set of outputs (\emph{e.g.} messages).
A code, then, implies a representation of the input variable. When representations are shared among entities, they become conventions which are used for
communication \citep{Burgos2014, Burgos2015}. The correct use of these conventions for communicating can be interpreted as ``honest signalling''.
For instance, the TCP/IP protocol allows the interaction of hardware and software in a code-based, ``plug-and-play'' fashion, as long as they obey
the protocol \citep{Doyle2010}. In biology, the genetic code acts as an innovation-sharing protocol, one that allows the exchange of innovations
through horizontal gene transfer (HGT) \citep{Woese2004}. 

However, communication protocols introduce vulnerabilities: parasitic agents can take advantage of them \citep{Ackley1994, Doyle2010}.
For instance, the chemical cues that ant colonies use to recognise nest-mates can be mimicked by slave-making workers for social integration
\citep{dEttorre2002}. On the Internet, one can take advantage of machine communication protocols (TCP/IP) to force target computers to
perform computations on behalf of a remote node, in what is called ``parasitic computing'' \citep{Barabasi2001}.

Parasites benefit from their interaction with other agents, while reducing the fitness of the attacked hosts. Nevertheless, parasites can be a
positive force in evolution. For instance, they can be generators of biodiversity, achieving more resistance to future attacks \citep{Brockhurst2004}.
In an artificial setting, the presence of parasites was shown to attain more efficient communication between agents of a population,
increasing their reproductive success \citep{Robbins1994}. Furthermore, some authors suggest that a healthy ecosystem is one rich in parasites
\citep{Hudson2006}.  In this work, we study this apparent contradiction from an information-theoretic perspective.

%For instance, they can manipulate their host to their advantage \citep{Schmid-Hempel1998}, they are
%believed to have influence in the evolution of sex \citep{Hamilton1990}, and they are generators of biodiversity \citep{Brockhurst2004}. 
%Nonetheless, the main characteristic of parasites is that they
%reduce the fitness of the attacked host. 

We look at some aspects of the co-evolutionary arms race between host and parasite. Particularly, we would like to characterise informationally the
behaviour of parasites and the consequences for the host. For this purpose, we assume a simple scenario where organisms seek to maximise their
long-term growth rate by following a bet-hedging strategy \citep{Seger1987}. We know that maximising their information about
the environment achieves this \citep{Shannon1948, Kelly1956}. Then, individuals obtaining extra environmental information from other individuals will
have an advantage over those that do not, since they would be able to better predict the future environmental conditions \citep{DonaldsonMatasci2010}.
However, as we showed in a previous work, for simple agents which do not have the ability to identify who they are listening to, a shared code
among the population is necessary to interpret the transmitted information and therefore improve predictions \citep{Burgos2014, Burgos2015}. Here,
we keep this assumption with respect to the agents, and we study the effects of introducing a parasite in a population that previously evolved its
codes as well as its structure.

\section{Model}
\label{sec:model}

In our previous model of code evolution \citep{Burgos2014, Burgos2015}, the outputs or messages of an agent were produced according to a code,
which was a conditional probability from sensor states to messages. The probability of each sensor state of an agent conditioned on the environmental
variable $\mu$ was given. The information about the environment of each agent was obtained by considering the mutual information between the
environmental variable and its sensor variable, together with the outputs of other agents. These outputs would be perceived or not, according to
the structure of the population. The codes, as well as the population structure, were optimised in order to maximise what was called the
similarity of the codes (we will introduce a more suitable term below) among the interacting agents of the population.

Here, instead, we consider a simplified model where the sensor states of an agent are the agent's messages, which are represented by a random variable
$X_\Theta$. That is, $p\left(X_\Theta \;\middle\vert\; \mu, \Theta = \theta\right)$ gives the probability distribution of the sensor states (and,
simultaneously, the messages) of an agent $\theta$ given the environmental conditions $\mu$.

% talk about roles here
Agents perceive the sensor states (messages) of other agents according to the structure of the population, which is given by $p(\Theta, \Theta^\prime)$.
This joint probability induces a weighted graph, where agents represent the nodes of the graph and there is an edge from agent $\theta$ to an agent
$\theta^\prime$ if $p(\theta, \theta^\prime) > 0$ (which is the weight of the edge). We interpret $p(\theta, \theta^\prime)$ as the probability
of interaction between these two agents, and thus we require that $p(\theta, \theta^\prime) = p(\theta^\prime, \theta)$ (interactions are
symmetrical) and $p(\theta, \theta) = 0$ for every agent $\theta$ (self-interaction is excluded).

\begin{figure}[ht]
\centering
  \begin{tikzpicture}
    [->,>=stealth',shorten >=2pt,auto,node distance=2cm,
    thick,main node/.style={font=\sffamily\normalsize\bfseries}]

    \node[main node] (1) [] {$\mu$};
    \node[main node] (2) [below left of=1] {$X_{\Theta}$};
    \node[main node] (3) [below right of=1] {$X_{\Theta^\prime}$};
    \node[main node] (4) [below of=2] {$\Theta$};
    \node[main node] (5) [below of=3] {$\Theta^\prime$};
    \node[main node] (6) [below of=1, node distance=5cm] {$\Xi$};

    \path[every node/.style={font=\sffamily\small}]
      (1) edge node {} (2)
      (1) edge node {} (3)
      (4) edge node {} (2)
      (5) edge node {} (3)
      (6) edge node {} (4)
      (6) edge node {} (5)
      ;
  \end{tikzpicture}
  \caption{\small{Bayesian network representing the relation of the variables in the simplified model of code evolution. $X_{\Theta^\prime}$ is an \emph{i.i.d}
		copy of $X_\Theta$. $\Theta$ and $\Theta^\prime$ selects agents from the same set, but their probability distributions are not necessary the same. These
		two variables depend on a common variable $\Xi$ to model more general interaction structures.}}
  \label{fig:model_simplified}
\end{figure}
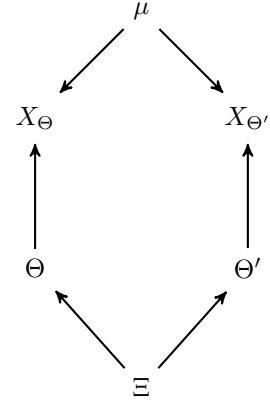

We now consider a population of agents where interacting agents maximise their \emph{mutual understanding}. This is formally
defined by $I\left(X_\Theta \;;\; X_{\Theta^\prime}\right)$, and, when this value achieves its maximum, the mapping that results from the agents'
codes, $p\left(X_\Theta \;\middle\vert\; X_{\Theta^\prime}\right)$, is deterministic. It is important to note here that this model allows
the agents to cluster into different sub-populations due to differences in their codes. Therefore, each sub-population could have its own
convention for representing different aspects about the environment, and the conventions used can be as varied as possible, as long as the mapping
$p\left(X_\Theta \;\middle\vert\; X_{\Theta^\prime}\right)$ is universal among all sub-populations.

% say when the mutual understanding is locally optimal
For cases where the mutual understanding is locally optimal, the codes of the agents are related to each other in one of two possible manners: (a) within
each sub-population, all agents have the same code, and the mapping $p\left(X_\Theta \;\middle\vert\; X_{\Theta^\prime}\right)$ is the identity matrix;
or (b) within each sub-population, there are two types of agents (where the type is given by the agent's code), and the interaction is only between
agents of different type. In this case, the graph induced by the interaction probability is bipartite between the types. %Then, each message $x_\Theta$
%maps to a message $x_{\Theta^\prime}$, where these two messages are different in at least two of all the resulting mapping between messages.

For cases where an agent interacts with more than one type of agent, then $p\left(X_\Theta \;\middle\vert\; X_{\Theta^\prime}\right)$
will necessarily be probabilistic, and thus the mutual understanding among the population will decrease. We can measure the total amount of information
about the environment of an agent $\theta$ by $I\left(\mu \;;\; X_\Theta, X_{\Theta^\prime} \;\middle\vert\; \Theta = \theta\right)$. And, since the
interaction probability is symmetric, the proposed measure for agent $\theta$ is equal to
$I\left(\mu \;;\; X_\Theta, X_{\Theta^\prime} \;\middle\vert\; \Theta^\prime = \theta\right)$.
Let us note that, whenever the mutual understanding of a population is optimal, then the individuals that interact necessarily capture the same
aspects of the environment. Then, at the optimum of mutual understanding in a population, agents do not increase their information by reading
other agent's messages, although this indeed plays an important role in the evolution of codes. Nevertheless, the ties that an agent establishes
are relevant for other purposes, which we study in the following sections.

\subsection{Informational parasitism}

There are different ways to define an informational parasite. Here, we adopt the model that characterises an informational parasite as an agent $\pi$
that tries to minimise the mutual understanding between the agents with whom it interacts. An informationally antagonistic parasite is not typical
for biology, as the parasite is concentrating at abusing the host system for its own interest, but does not care about the host except for avoiding
detection. However, in the context of social networks or news sources, such a parasite can be considered a ``troll'' or a ``disinformation'' (FUD)
agent who has direct interest in damaging the mutual understanding of the other agents of the population and/or their confidence in their knowledge
of the true state of the environment.

In our case, the parasite will choose its code in such a way that the value $I\left(X_\Theta \;;\; X_{\Theta^\prime}\right)$ is minimised. 
This is an extreme case of parasitism, where the parasite may kill its host as a result of maximising damage. Usually, the known parasites manipulate
their hosts in order to benefit from it, decreasing their fitness such that it would not kill them \citep{Schmid-Hempel1998}. Although the defined
type of parasite is not common in biology, it is still a possibility in the range of behaviours that decrease fitness of the host while increasing
the attacker's fitness. %We expect that, as a consequence, the parasite will maximise its information about the environment. 

We now analyse the consequences of introducing a parasite in a population for a few very simple, but illustrative, scenarios. First, let us define
the following types of codes:

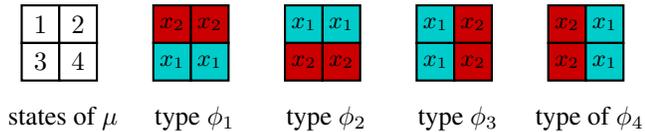
\begin{figure}[ht]
	\centering
	\begin{tikzpicture}
		\def\bc{0.75} % space between codes
		\def\as{1} % added space for each code
		% column nodes
		%\node[] at (0.125, 0.5) {\tiny{$x_1$}};
		%\node[] at (0.375, 0.5) {\tiny{$x_2$}};
		%\node[] at (0.625, 0.5) {\tiny{$x_3$}};
		%\node[] at (0.875, 0.5) {\tiny{$x_4$}};
		%\draw[decorate,decoration={brace,mirror,raise=6pt,amplitude=10pt}, thick] (0 0)--(0 1) 
		%\node[] at (-0.3, -0.125) {\footnotesize{$y_1$}};
		%\node[] at (-0.3, -0.375) {\footnotesize{$y_2$}};
		%\node[] at (-0.3, -0.625) {\footnotesize{$y_3$}};
		%\node[] at (-0.3, -0.875) {\footnotesize{$y_4$}};
		% code number
		\def\code{0}
		\filldraw[draw=black,thick,fill={rgb,1:red,1;green,1;blue,1}] (0+\code+\code*\bc,-0) rectangle (0.5+\code+\code*\bc,-0.5);
		\filldraw[draw=black,thick,fill={rgb,1:red,1;green,1;blue,1}] (0.5+\code+\code*\bc,-0) rectangle (1.0+\code+\code*\bc,-0.5);
		\filldraw[draw=black,thick,fill={rgb,1:red,1;green,1;blue,1}] (0+\code+\code*\bc,-0.5) rectangle (0.5+\code+\code*\bc,-1.0);
		\filldraw[draw=black,thick,fill={rgb,1:red,1;green,1;blue,1}] (0.5+\code+\code*\bc,-0.5) rectangle (1.0+\code+\code*\bc,-1.0);
		\node[] at (0.25+\code+\code*\bc, -0.25) {$1$};
		\node[] at (0.75+\code+\code*\bc, -0.25) {$2$};
		\node[] at (0.25+\code+\code*\bc, -0.75) {$3$};
		\node[] at (0.75+\code+\code*\bc, -0.75) {$4$};
		\node[] at (0.55+\code+\code*\bc, -1.5) {states of $\mu$};
		% code number
		\def\code{1}
		\filldraw[draw=black,thick,fill={rgb,1:red,0.8;green,0;blue,0}] (0+\code+\code*\bc,-0) rectangle (0.5+\code+\code*\bc,-0.5);
		\filldraw[draw=black,thick,fill={rgb,1:red,0.8;green,0;blue,0}] (0.5+\code+\code*\bc,-0) rectangle (1.0+\code+\code*\bc,-0.5);
		\filldraw[draw=black,thick,fill={rgb,1:red,0;green,0.8;blue,0.8}] (0+\code+\code*\bc,-0.5) rectangle (0.5+\code+\code*\bc,-1.0);
		\filldraw[draw=black,thick,fill={rgb,1:red,0;green,0.8;blue,0.8}] (0.5+\code+\code*\bc,-0.5) rectangle (1.0+\code+\code*\bc,-1.0);
		\node[] at (0.25+\code+\code*\bc, -0.25) {\small{$x_2$}};
		\node[] at (0.75+\code+\code*\bc, -0.25) {\small{$x_2$}};
		\node[] at (0.25+\code+\code*\bc, -0.75) {\small{$x_1$}};
		\node[] at (0.75+\code+\code*\bc, -0.75) {\small{$x_1$}};
		\node[] at (0.55+\code+\code*\bc, -1.5) {type $\phi_1$};
		% code number
		\def\code{2}
		\filldraw[draw=black,thick,fill={rgb,1:red,0;green,0.8;blue,0.8}] (0+\code+\code*\bc,-0) rectangle (0.5+\code+\code*\bc,-0.5);
		\filldraw[draw=black,thick,fill={rgb,1:red,0;green,0.8;blue,0.8}] (0.5+\code+\code*\bc,-0) rectangle (1.0+\code+\code*\bc,-0.5);
		\filldraw[draw=black,thick,fill={rgb,1:red,0.8;green,0;blue,0}] (0+\code+\code*\bc,-0.5) rectangle (0.5+\code+\code*\bc,-1.0);
		\filldraw[draw=black,thick,fill={rgb,1:red,0.8;green,0;blue,0}] (0.5+\code+\code*\bc,-0.5) rectangle (1.0+\code+\code*\bc,-1.0);
		\node[] at (0.25+\code+\code*\bc, -0.25) {\small{$x_1$}};
		\node[] at (0.75+\code+\code*\bc, -0.25) {\small{$x_1$}};
		\node[] at (0.25+\code+\code*\bc, -0.75) {\small{$x_2$}};
		\node[] at (0.75+\code+\code*\bc, -0.75) {\small{$x_2$}};
		\node[] at (0.55+\code+\code*\bc, -1.5) {type $\phi_2$};
		% code number
		\def\code{3}
		\filldraw[draw=black,thick,fill={rgb,1:red,0;green,0.8;blue,0.8}] (0+\code+\code*\bc,-0) rectangle (0.5+\code+\code*\bc,-0.5);
		\filldraw[draw=black,thick,fill={rgb,1:red,0.8;green,0;blue,0}] (0.5+\code+\code*\bc,-0) rectangle (1.0+\code+\code*\bc,-0.5);
		\filldraw[draw=black,thick,fill={rgb,1:red,0;green,0.8;blue,0.8}] (0+\code+\code*\bc,-0.5) rectangle (0.5+\code+\code*\bc,-1.0);
		\filldraw[draw=black,thick,fill={rgb,1:red,0.8;green,0;blue,0}] (0.5+\code+\code*\bc,-0.5) rectangle (1.0+\code+\code*\bc,-1.0);
		\node[] at (0.25+\code+\code*\bc, -0.25) {\small{$x_1$}};
		\node[] at (0.75+\code+\code*\bc, -0.25) {\small{$x_2$}};
		\node[] at (0.25+\code+\code*\bc, -0.75) {\small{$x_1$}};
		\node[] at (0.75+\code+\code*\bc, -0.75) {\small{$x_2$}};
		\node[] at (0.55+\code+\code*\bc, -1.5) {type $\phi_3$};
		% code number
		\def\code{4}
		\filldraw[draw=black,thick,fill={rgb,1:red,0.8;green,0;blue,0}] (0+\code+\code*\bc,-0) rectangle (0.5+\code+\code*\bc,-0.5);
		\filldraw[draw=black,thick,fill={rgb,1:red,0;green,0.8;blue,0.8}] (0.5+\code+\code*\bc,-0) rectangle (1.0+\code+\code*\bc,-0.5);
		\filldraw[draw=black,thick,fill={rgb,1:red,0.8;green,0;blue,0}] (0+\code+\code*\bc,-0.5) rectangle (0.5+\code+\code*\bc,-1.0);
		\filldraw[draw=black,thick,fill={rgb,1:red,0;green,0.8;blue,0.8}] (0.5+\code+\code*\bc,-0.5) rectangle (1.0+\code+\code*\bc,-1.0);
		\node[] at (0.25+\code+\code*\bc, -0.25) {\small{$x_2$}};
		\node[] at (0.75+\code+\code*\bc, -0.25) {\small{$x_1$}};
		\node[] at (0.25+\code+\code*\bc, -0.75) {\small{$x_2$}};
		\node[] at (0.75+\code+\code*\bc, -0.75) {\small{$x_1$}};
		\node[] at (0.55+\code+\code*\bc, -1.5) {type of $\phi_4$};
	\end{tikzpicture}
	\caption{\small{The left-most grid shows an illustration of the environment, although it does not denote its real structure. Then, each type shows a
			partition of the environmental states induced by an agent's code (codes here are deterministic). The types $\phi_1$ and $\phi_2$ capture the first bit of $\mu$.
			Types $\phi_3$ and $\phi_4$ capture the second bit of $\mu$.}}
  \label{fig:code_intro}
\end{figure}

%\begin{equation}
%	p_1\left(x \;\middle\vert\; \mu\right) \coloneqq 
%	\bordermatrix{~ & x_1 & x_2 \cr
%	              \mu_1 & 0 & 1 \cr
%	              \mu_2 & 0 & 1 \cr
%	              \mu_3 & 1 & 0 \cr
%	              \mu_4 & 1 & 0 \cr}
%	\label{eq:cp1}
%\end{equation}
%\begin{equation}
%	p_2\left(x \;\middle\vert\; \mu\right) \coloneqq 
%	\bordermatrix{~ & x_1 & x_2 \cr
%	              \mu_1 & 1 & 0 \cr
%	              \mu_2 & 1 & 0 \cr
%	              \mu_3 & 0 & 1 \cr
%	              \mu_4 & 0 & 1 \cr}
%	\label{eq:cp2}
%\end{equation}
%\begin{equation}
%	p_3\left(x \;\middle\vert\; \mu\right) \coloneqq 
%	\bordermatrix{~ & x_1 & x_2 \cr
%	              \mu_1 & 1 & 0 \cr
%	              \mu_2 & 0 & 1 \cr
%	              \mu_3 & 1 & 0 \cr
%	              \mu_4 & 0 & 1 \cr}
%	\label{eq:cp3}
%\end{equation}

Let us analyse a few simple scenarios where a parasite attacks a population. Let us assume that two (non-parasitic) agents share the same code,
for instance, agents $\theta_1$ and $\theta_2$ are of type $\phi_1$, and that these agents interact only with each other (their mutual understanding
is $1$ bit). Now, if we minimise their mutual understanding by introducing a parasite $\pi$, the optimal configuration is the following: the parasite
interacts with one agent only and the parasite's code is of type $\phi_2$ (the ``opposite'' of type $\phi_1$). The mutual understanding between all
three agents now is zero. Let us note that, in this case, the environmental information of each agent, before and after introducing the parasite,
is $1$ bit, which is the amount of information each of them acquire through their corresponding sensors.

%Let us consider now a more interesting case: two sub-populations of two agents each, where agents only interact with agents within the sub-population
%and where the codes are the following: in the first sub-population, we have agents $\theta_1$ and $\theta_2$ of code type $\phi_1$; in the second
%sub-population, we have agents $\theta_3$ and $\theta_4$ of code type $\phi_2$. That is, both sub-populations capture the same aspects of $\mu$, but
%they are expressed using different conventions if they are considered altogether. Since the conventions used by the sub-populations are ``opposite'',
%then, in this case, the parasite cannot interact with both sub-populations and decrease their mutual understanding simultaneously. Here, the minimum
%value of $I\left(X_\Theta \;;\; X_{\Theta^\prime}\right) = 0$ bits is achieved with two configurations: in the first one, the parasite adopts the code
%type of $\theta_1$ and $\theta_2$ (which is the opposite code of agents $\theta_3$ and $\theta_4$), and interacts \emph{only} with agents $\theta_3$
%and $\theta_4$; and, in the second case, the parasite adopts the code type of $\theta_3$ and $\theta_4$ and interacts \emph{only} with agents $\theta_1$
%and $\theta_2$. As before, the environmental information of each agent remains equal to $1$ bit, which is what they acquire through their sensors.

Let us consider now a more interesting case: two sub-populations of two agents each, where agents only interact with agents within the sub-population
and where the codes are the following: in the first sub-population, we have agents $\theta_1$ and $\theta_2$ of code type $\phi_1$ (they capture the
first bit of $\mu$); and in the second sub-population, we have agents $\theta_3$ and $\theta_4$ of code type $\phi_3$ (they capture the second bit
of $\mu$). 
%Let us consider a similar case, but this time each sub-population capture different aspects of the environment: the first sub-population is composed
%of agents $\theta_1$ and $\theta_2$, which are of code type $\phi_1$ (they capture the first bit of $\mu$); and the second sub-population is composed
%of agents $\theta_3$ and $\theta_4$, which are of code type $\phi_3$ (they capture the second bit of $\mu$). 
Now, when we introduce a parasite, two
configurations achieve zero mutual understanding: in both, the parasite interacts with every agent, but in the first one, it adopts the ``opposite'' code
of agents $\theta_1$ and $\theta_2$, which is code type $\phi_2$, while in the second one, it adopts the ``opposite'' code of agents $\theta_3$ and
$\theta_4$, which is code type $\phi_4$. Here, in the first case, the environmental information of agents $\theta_3$ and $\theta_4$ increases, since
the parasite conveys information captured by agents $\theta_1$ and $\theta_2$ (captured by the parasite through its ``opposite'' code) that the former
two agents do not possess. In the same way, for the second configuration, agents $\theta_1$ and $\theta_2$ are benefited by their interactions with
the parasite, since here also, the parasite conveys information they lack.

%Finally, let us assume two sub-populations where the first captures the first bit of $\mu$, using $x \in \{ 1, 2 \}$, and the second captures
%the second bit of $\mu$, using $x \in \{ 3, 4 \}$, that is, non-overlapping messages. After the introduction of the parasite, the population's
%mutual understanding is $I\left(X_\Theta \;;\; X_{\Theta^\prime}\right) = 0.0817$ bits, and the code of the parasite is such that it captures
%all information about $\mu$. In this case all agents improve their environmental information, and the parasite can perfectly predict the state of the
%environment.

\section{Methods}
\label{sec:methods}

%In order to accelerate computations, we made three assumptions about the model: first, the probability of the environmental states is
%uniformly distributed; second, the codes of the agents is deterministic given the environment; and third, the interaction probability
%is uniform among those agents that interact.

All optimisations were performed using a genetic algorithm (GA). We utilised the C++ library GAlib v2.4.7 \citep{Wall1996}. The
GA searches for a particular (possibly local) optimum, and this optimum corresponds to an evolutionary process that has converged.
In order to accelerate computations, we assume the following: the probability distribution over $\mu$ is uniform, the codes of all
agents are deterministic, and the interaction probability between any two agents is given by one over the amount of interactions.
%In the path from the initial configuration to an optimum, the possible changes in code and structure are not...
%The parameters used were the following: populations per generation: $50$; probability of mutating the structure: $0.35$; probability
%of mutating a code: $0.6$. No genetic crossings were allowed.
To visualise the evolution of the codes of the agents, we use the method of multidimensional scaling provided by R version 2.14.1
(2011-12-22). This method takes as input the distance matrix between codes, and plots them in a two-dimensional space preserving the distances
as well as possible. 

\section{Results}
\label{sec:results}

% it attack the symbols present in most codes, show p(x0,x1)
% code distance is log2 of the amount of different symbols for expressing the same conditions

\begin{figure}[ht]
	\centering
	\includegraphics[scale=0.5]{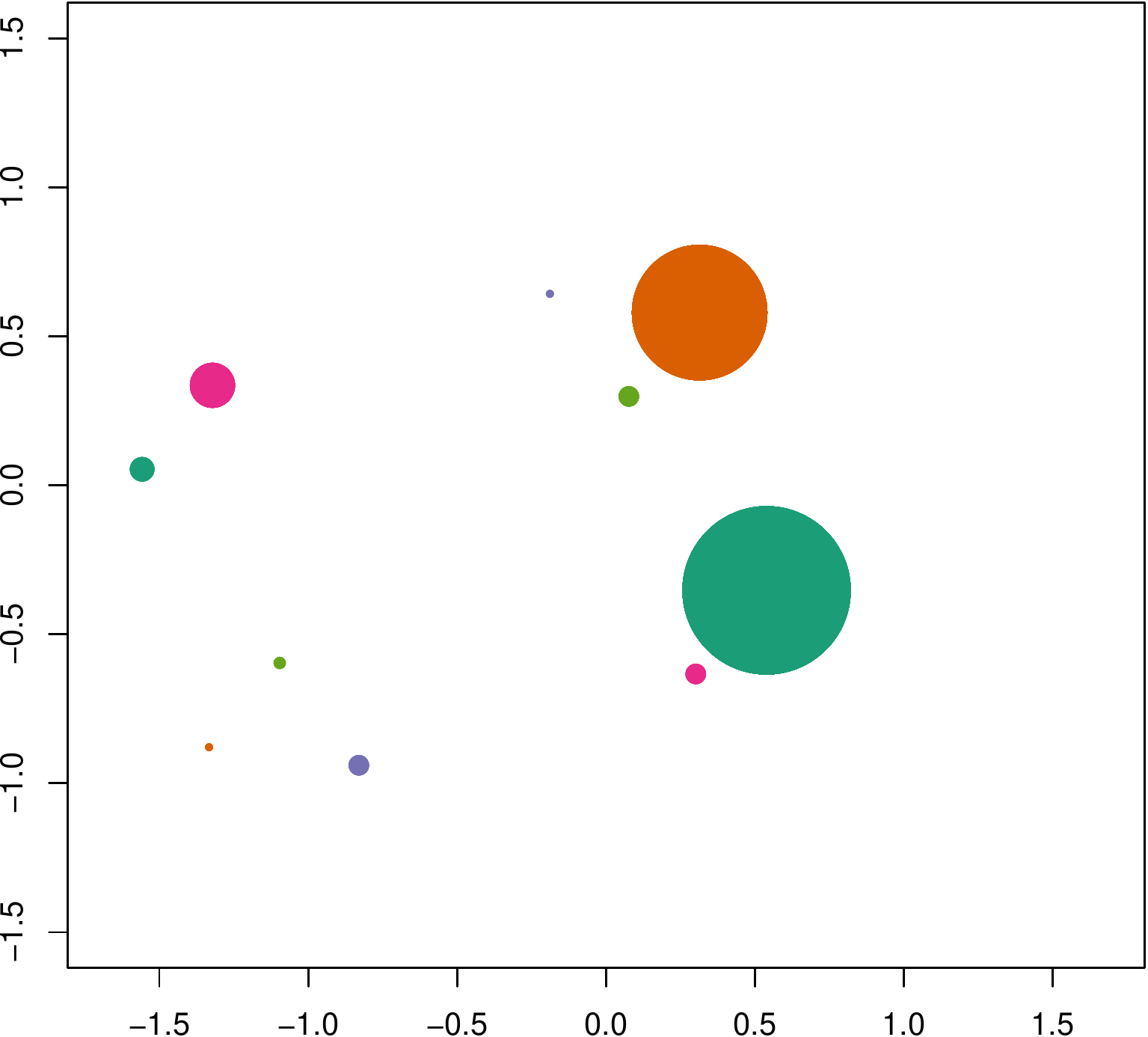}
	\caption{\small{2-dimensional plot of the distance between the codes: each point represents a particular code, and its size is relative to
			the number of agents adopting that particular code. The colour of the points denotes the sub-population to which the codes belong.}}
	\label{fig:pardistplot}
\end{figure}

We study the introduction of a parasite in a population where the mutual understanding was previously maximised. We consider a
population of $256$ agents, in an environment with $16$ equally likely states, where agents can encode the environment using $16$ different
symbols. In this way, agents can potentially capture by themselves all the information about $\mu$. As a result of the optimisation process,
we obtained $5$ sub-populations, where, in each of them, the induced interaction graph is bipartite. Therefore, there are $10$ different codes
globally, two per sub-population. In Fig. \ref{fig:pardistplot} we show the distance between the resulting codes, with point's size proportional
to the number of agents that adopted each code. The distance used is the Jensen-Shannon divergence (see \citep{Burgos2014}). The average mutual
understanding in the population is $I\left(X_\Theta \;;\; X_{\Theta^\prime}\right) = 3.93$ bits, which, coincidentally, is also the mutual
understanding within each sub-population (this does not need to hold necessarily).

% should I show if the sub-populations capture similar/different aspects?
% when does a population start getting damaged by the parasite (in terms of environmental information)?
% should I plot how identifiable the parasites are?

\subsection{Blending in with the crowd}
\label{sec:blendin}

% mention counterfeit signals (if possible)
We introduce now a parasite $\pi$ in the population, and we let it freely choose with whom it interacts, as well as its code (the parasite
is introduced before the optimisation process begins, at generation $0$). However, the
parasite is allowed to use $32$ symbols to encode $\mu$, instead of $16$, as we did for the rest of the agents. The reason for allowing
a larger set of symbols to the parasite is that, otherwise, we will be forcing the parasite to use the symbols used by the population. We allow
the double amount of symbols to enable the parasite to perfectly encode the environment while avoiding all symbols already in use. 

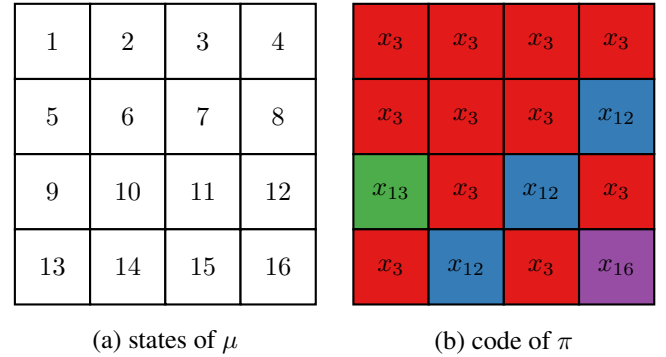
\begin{figure}[ht]
	\centering
	\begin{tikzpicture}
		\def\bc{3.5} % space between codes
		\def\as{0} % added space for each code
		% code number
		\def\code{0}
		\filldraw[draw=black,thick,fill={rgb,1:red,1;green,1;blue,1}] (0+\code+\code*\bc,-0) rectangle (1.0+\code+\code*\bc,-1.0);
		\filldraw[draw=black,thick,fill={rgb,1:red,1;green,1;blue,1}] (1.0+\code+\code*\bc,-0) rectangle (2.0+\code+\code*\bc,-1.0);
		\filldraw[draw=black,thick,fill={rgb,1:red,1;green,1;blue,1}] (2.0+\code+\code*\bc,-0) rectangle (3.0+\code+\code*\bc,-1.0);
		\filldraw[draw=black,thick,fill={rgb,1:red,1;green,1;blue,1}] (3.0+\code+\code*\bc,-0) rectangle (4.0+\code+\code*\bc,-1.0);
		\filldraw[draw=black,thick,fill={rgb,1:red,1;green,1;blue,1}] (0+\code+\code*\bc,-1.0) rectangle (1.0+\code+\code*\bc,-2.0);
		\filldraw[draw=black,thick,fill={rgb,1:red,1;green,1;blue,1}] (1.0+\code+\code*\bc,-1.0) rectangle (2.0+\code+\code*\bc,-2.0);
		\filldraw[draw=black,thick,fill={rgb,1:red,1;green,1;blue,1}] (2.0+\code+\code*\bc,-1.0) rectangle (3.0+\code+\code*\bc,-2.0);
		\filldraw[draw=black,thick,fill={rgb,1:red,1;green,1;blue,1}] (3.0+\code+\code*\bc,-1.0) rectangle (4.0+\code+\code*\bc,-2.0);
		\filldraw[draw=black,thick,fill={rgb,1:red,1;green,1;blue,1}] (0+\code+\code*\bc,-2.0) rectangle (1.0+\code+\code*\bc,-3.0);
		\filldraw[draw=black,thick,fill={rgb,1:red,1;green,1;blue,1}] (1.0+\code+\code*\bc,-2.0) rectangle (2.0+\code+\code*\bc,-3.0);
		\filldraw[draw=black,thick,fill={rgb,1:red,1;green,1;blue,1}] (2.0+\code+\code*\bc,-2.0) rectangle (3.0+\code+\code*\bc,-3.0);
		\filldraw[draw=black,thick,fill={rgb,1:red,1;green,1;blue,1}] (3.0+\code+\code*\bc,-2.0) rectangle (4.0+\code+\code*\bc,-3.0);
		\filldraw[draw=black,thick,fill={rgb,1:red,1;green,1;blue,1}] (0+\code+\code*\bc,-3.0) rectangle (1.0+\code+\code*\bc,-4.0);
		\filldraw[draw=black,thick,fill={rgb,1:red,1;green,1;blue,1}] (1.0+\code+\code*\bc,-3.0) rectangle (2.0+\code+\code*\bc,-4.0);
		\filldraw[draw=black,thick,fill={rgb,1:red,1;green,1;blue,1}] (2.0+\code+\code*\bc,-3.0) rectangle (3.0+\code+\code*\bc,-4.0);
		\filldraw[draw=black,thick,fill={rgb,1:red,1;green,1;blue,1}] (3.0+\code+\code*\bc,-3.0) rectangle (4.0+\code+\code*\bc,-4.0);
		\node[] at (0.5+\code+\code*\bc, -0.5) {$1$};
		\node[] at (1.5+\code+\code*\bc, -0.5) {$2$};
		\node[] at (2.5+\code+\code*\bc, -0.5) {$3$};
		\node[] at (3.5+\code+\code*\bc, -0.5) {$4$};
		\node[] at (0.5+\code+\code*\bc, -1.5) {$5$};
		\node[] at (1.5+\code+\code*\bc, -1.5) {$6$};
		\node[] at (2.5+\code+\code*\bc, -1.5) {$7$};
		\node[] at (3.5+\code+\code*\bc, -1.5) {$8$};
		\node[] at (0.5+\code+\code*\bc, -2.5) {$9$};
		\node[] at (1.5+\code+\code*\bc, -2.5) {$10$};
		\node[] at (2.5+\code+\code*\bc, -2.5) {$11$};
		\node[] at (3.5+\code+\code*\bc, -2.5) {$12$};
		\node[] at (0.5+\code+\code*\bc, -3.5) {$13$};
		\node[] at (1.5+\code+\code*\bc, -3.5) {$14$};
		\node[] at (2.5+\code+\code*\bc, -3.5) {$15$};
		\node[] at (3.5+\code+\code*\bc, -3.5) {$16$};
		\node[] at (2.0+\code+\code*\bc, -4.5) {(a) states of $\mu$};
		% code number
		\def\code{1}
		\filldraw[draw=black,thick,fill={rgb,1:red,0.89;green,0.1;blue,0.1}] (0+\code+\code*\bc,-0) rectangle (1.0+\code+\code*\bc,-1.0);
		\filldraw[draw=black,thick,fill={rgb,1:red,0.89;green,0.1;blue,0.1}] (1.0+\code+\code*\bc,-0) rectangle (2.0+\code+\code*\bc,-1.0);
		\filldraw[draw=black,thick,fill={rgb,1:red,0.89;green,0.1;blue,0.1}] (2.0+\code+\code*\bc,-0) rectangle (3.0+\code+\code*\bc,-1.0);
		\filldraw[draw=black,thick,fill={rgb,1:red,0.89;green,0.1;blue,0.1}] (3.0+\code+\code*\bc,-0) rectangle (4.0+\code+\code*\bc,-1.0);
		\filldraw[draw=black,thick,fill={rgb,1:red,0.89;green,0.1;blue,0.1}] (0+\code+\code*\bc,-1.0) rectangle (1.0+\code+\code*\bc,-2.0);
		\filldraw[draw=black,thick,fill={rgb,1:red,0.89;green,0.1;blue,0.1}] (1.0+\code+\code*\bc,-1.0) rectangle (2.0+\code+\code*\bc,-2.0);
		\filldraw[draw=black,thick,fill={rgb,1:red,0.89;green,0.1;blue,0.1}] (2.0+\code+\code*\bc,-1.0) rectangle (3.0+\code+\code*\bc,-2.0);
		\filldraw[draw=black,thick,fill={rgb,1:red,0.21;green,0.49;blue,0.72}] (3.0+\code+\code*\bc,-1.0) rectangle (4.0+\code+\code*\bc,-2.0);
		\filldraw[draw=black,thick,fill={rgb,1:red,0.30;green,0.68;blue,0.29}] (0+\code+\code*\bc,-2.0) rectangle (1.0+\code+\code*\bc,-3.0);
		\filldraw[draw=black,thick,fill={rgb,1:red,0.89;green,0.1;blue,0.1}] (1.0+\code+\code*\bc,-2.0) rectangle (2.0+\code+\code*\bc,-3.0);
		\filldraw[draw=black,thick,fill={rgb,1:red,0.21;green,0.49;blue,0.72}] (2.0+\code+\code*\bc,-2.0) rectangle (3.0+\code+\code*\bc,-3.0);
		\filldraw[draw=black,thick,fill={rgb,1:red,0.89;green,0.1;blue,0.1}] (3.0+\code+\code*\bc,-2.0) rectangle (4.0+\code+\code*\bc,-3.0);
		\filldraw[draw=black,thick,fill={rgb,1:red,0.89;green,0.1;blue,0.1}] (0+\code+\code*\bc,-3.0) rectangle (1.0+\code+\code*\bc,-4.0);
		\filldraw[draw=black,thick,fill={rgb,1:red,0.21;green,0.49;blue,0.72}] (1.0+\code+\code*\bc,-3.0) rectangle (2.0+\code+\code*\bc,-4.0);
		\filldraw[draw=black,thick,fill={rgb,1:red,0.89;green,0.1;blue,0.1}] (2.0+\code+\code*\bc,-3.0) rectangle (3.0+\code+\code*\bc,-4.0);
		\filldraw[draw=black,thick,fill={rgb,1:red,0.59;green,0.30;blue,0.64}] (3.0+\code+\code*\bc,-3.0) rectangle (4.0+\code+\code*\bc,-4.0);
		\node[] at (0.5+\code+\code*\bc, -0.5) {$x_3$};
		\node[] at (1.5+\code+\code*\bc, -0.5) {$x_3$};
		\node[] at (2.5+\code+\code*\bc, -0.5) {$x_3$};
		\node[] at (3.5+\code+\code*\bc, -0.5) {$x_3$};
		\node[] at (0.5+\code+\code*\bc, -1.5) {$x_3$};
		\node[] at (1.5+\code+\code*\bc, -1.5) {$x_3$};
		\node[] at (2.5+\code+\code*\bc, -1.5) {$x_3$};
		\node[] at (3.5+\code+\code*\bc, -1.5) {$x_{12}$};
		\node[] at (0.5+\code+\code*\bc, -2.5) {$x_{13}$};
		\node[] at (1.5+\code+\code*\bc, -2.5) {$x_3$};
		\node[] at (2.5+\code+\code*\bc, -2.5) {$x_{12}$};
		\node[] at (3.5+\code+\code*\bc, -2.5) {$x_3$};
		\node[] at (0.5+\code+\code*\bc, -3.5) {$x_3$};
		\node[] at (1.5+\code+\code*\bc, -3.5) {$x_{12}$};
		\node[] at (2.5+\code+\code*\bc, -3.5) {$x_3$};
		\node[] at (3.5+\code+\code*\bc, -3.5) {$x_{16}$};
		\node[] at (2.0+\code+\code*\bc, -4.5) {(b) code of $\pi$};
	\end{tikzpicture}
	%\caption{\small{Representation of the conditional probabilities $p(Y_\theta|\mu)$ for an agent $\theta$ of each type. These are defined such that
    %         each type of agent can only distinguish between the coloured region and the white region. For instance, the sensor of type $\phi_2$ is defined
    %         as $Pr(Y = y_1 | \mu) = 1$ if $\mu \in \{ 1, 2, 4, 5 \}$, and zero otherwise, and $Pr(Y = y_2 | \mu) = 1$ if $\mu \notin \{1, 2, 4, 5 \}$,
    %         and zero otherwise. For type $\phi_1$, $Pr(Y = y_1 | \mu) = 0.5$ and $Pr(Y = y_2 | \mu) = 0.5$ ($|Y| = 2$ for all types of agents).}}
%and the probability of each state is uniformly distributed. We illustrate this environment by a $3 \times 3$ grid, as shown in
%Fig. \ref{fig:sensor_types}, although the square does not denote the real structure of the environment. Then, the outputs of each type of agents will
	\caption{\small{(a) Illustration of the environment $\mu$, although the grid does not denote its real structure. (b) Partition of the environmental states
			induced by the code of the parasite $\pi$.}}
  \label{fig:code_types}
\end{figure}

We found that after optimisation, the parasite interacts with every agent of the population, and its code distance to every other agent is maximal
(the distance is $4$, the maximum achievable with $16$ states). The resulting average mutual understanding now is
$I\left(X_\Theta \;;\; X_{\Theta^\prime}\right) = 2.55$ bits (before the attack, it was $3.93$ bits), and the code of the parasite is shown
in Fig. \ref{fig:code_types}.

\begin{figure}[ht]
	\centering
	\includegraphics[scale=0.5]{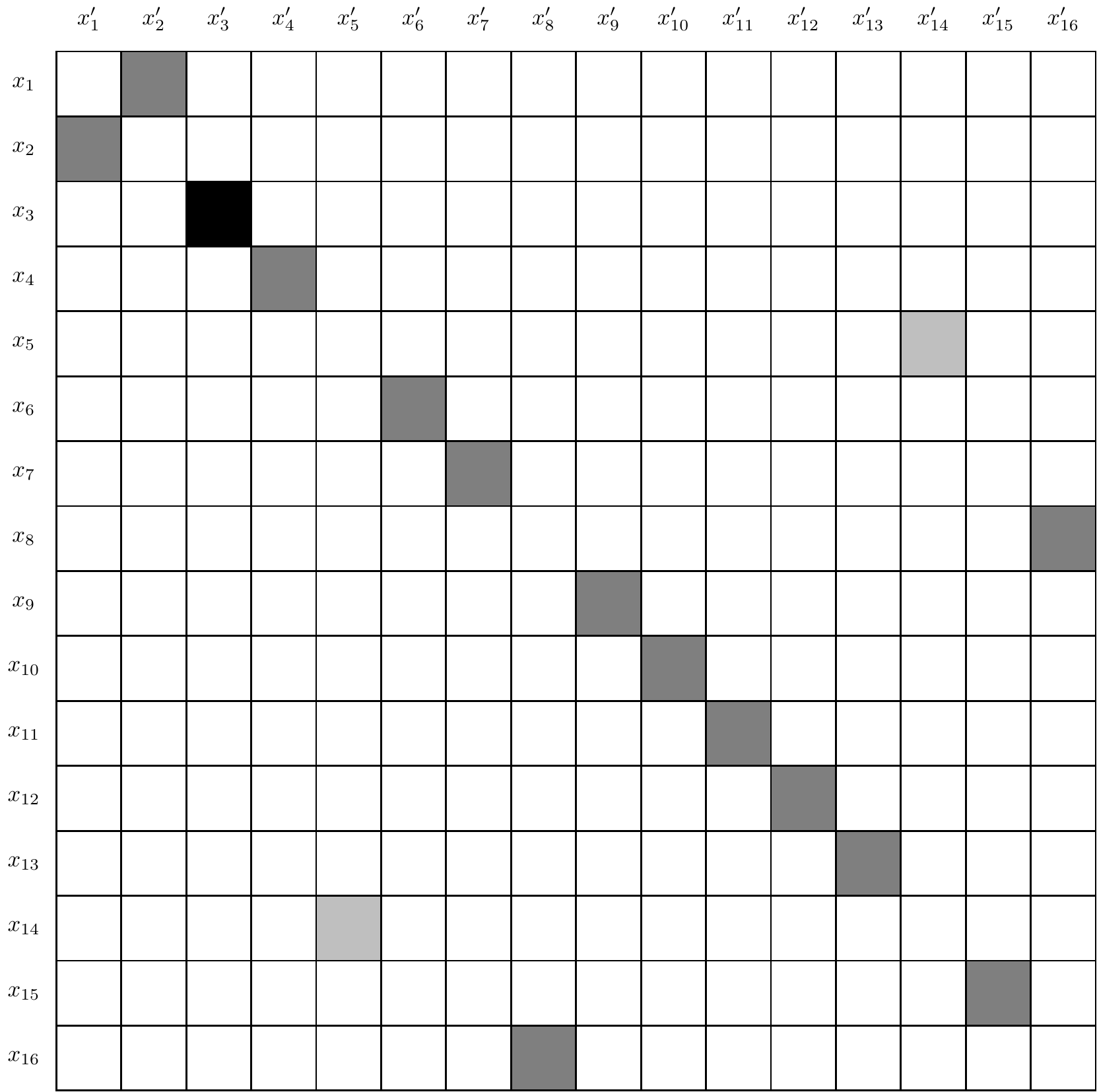}
	\caption{\small{Joint probability of messages $p(X_\Theta, X_{\Theta^\prime})$ of the population before the parasitic attack. Values are normalised to
			the maximum of all values, in this case $p(x_3, x^\prime_3) = 0.125$. White squares have probability zero.}}
	\label{fig:jointprobinit}
\end{figure}

To understand the choice of code by the parasite, we analyse the joint probability $p(X_\Theta, X_{\Theta^\prime})$ before introducing the
parasite in the population. Our results show that the parasite encodes the environment through the messages that are most commonly used among
the population. In this case, the parasite chose $4$ messages ($x_3$, $x_{12}$, $x_{13}$ and $x_{16}$), all of them among the most popular in
the population (see Fig. \ref{fig:jointprobinit}).

As a consequence of this antagonistic behaviour, the parasite blends in the population. This suggests that the parasite would try to avoid
being identified by its messages. Our model allows us to measure how ``identifiable'' agents are by comparing the average joint messages.
For instance, this can be measured by the mutual information between the agent selector and the joint messages,
$I\left(\Theta \;\middle\vert\; X_\Theta, X_{\Theta^\prime}\right)$. For a population with a universal code, this measure is zero, that is,
we cannot identify any agent. Here, we want to know particularly how much we can identify the parasite by its messages. Then, we can consider
the following measure:
%Then, we can consider the Kullback-Leibler divergence between the average messages and the average messages when the parsite is selected. Formaly,

\begin{equation}
D_{KL}\left( p(X_\Theta, X_{\Theta^\prime} \;\vert\; \Theta = \pi) \;\middle\vert\vert\; p(X_\Theta, X_{\Theta^\prime})\right)
\label{eq:KLdiv}
\end{equation}

\begin{figure}[ht]
	\centering
	\begin{tikzpicture}[gnuplot, scale = 0.6]
	%% generated with GNUPLOT 4.4p3 (Lua 5.1.4; terminal rev. 97, script rev. 96a)
	%% Tue 17 Mar 2015 04:20:35 GMT
	\gpcolor{gp lt color border}
	\gpsetlinetype{gp lt border}
	\gpsetlinewidth{1.00}
	\draw[gp path] (1.504,0.985)--(1.684,0.985);
	\draw[gp path] (11.947,0.985)--(11.767,0.985);
	\node[gp node right] at (1.320,0.985) {\small{2.9}};
	\draw[gp path] (1.504,2.042)--(1.684,2.042);
	\draw[gp path] (11.947,2.042)--(11.767,2.042);
	\node[gp node right] at (1.320,2.042) {\small{3}};
	\draw[gp path] (1.504,3.098)--(1.684,3.098);
	\draw[gp path] (11.947,3.098)--(11.767,3.098);
	\node[gp node right] at (1.320,3.098) {\small{3.1}};
	\draw[gp path] (1.504,4.155)--(1.684,4.155);
	\draw[gp path] (11.947,4.155)--(11.767,4.155);
	\node[gp node right] at (1.320,4.155) {\small{3.2}};
	\draw[gp path] (1.504,5.211)--(1.684,5.211);
	\draw[gp path] (11.947,5.211)--(11.767,5.211);
	\node[gp node right] at (1.320,5.211) {\small{3.3}};
	\draw[gp path] (1.504,6.268)--(1.684,6.268);
	\draw[gp path] (11.947,6.268)--(11.767,6.268);
	\node[gp node right] at (1.320,6.268) {\small{3.4}};
	\draw[gp path] (1.504,7.324)--(1.684,7.324);
	\draw[gp path] (11.947,7.324)--(11.767,7.324);
	\node[gp node right] at (1.320,7.324) {\small{3.5}};
	\draw[gp path] (1.504,8.381)--(1.684,8.381);
	\draw[gp path] (11.947,8.381)--(11.767,8.381);
	\node[gp node right] at (1.320,8.381) {\small{3.6}};
	\draw[gp path] (1.504,0.985)--(1.504,1.165);
	\draw[gp path] (1.504,8.381)--(1.504,8.201);
	\node[gp node center] at (1.504,0.477) {\small{0}};
	\draw[gp path] (3.593,0.985)--(3.593,1.165);
	\draw[gp path] (3.593,8.381)--(3.593,8.201);
	\node[gp node center] at (3.593,0.477) {\small{10}};
	\draw[gp path] (5.681,0.985)--(5.681,1.165);
	\draw[gp path] (5.681,8.381)--(5.681,8.201);
	\node[gp node center] at (5.681,0.477) {\small{20}};
	\draw[gp path] (7.770,0.985)--(7.770,1.165);
	\draw[gp path] (7.770,8.381)--(7.770,8.201);
	\node[gp node center] at (7.770,0.477) {\small{30}};
	\draw[gp path] (9.858,0.985)--(9.858,1.165);
	\draw[gp path] (9.858,8.381)--(9.858,8.201);
	\node[gp node center] at (9.858,0.477) {\small{40}};
	\draw[gp path] (11.947,0.985)--(11.947,1.165);
	\draw[gp path] (11.947,8.381)--(11.947,8.201);
	\node[gp node center] at (11.947,0.477) {\small{50}};
	\draw[gp path] (1.504,8.381)--(1.504,0.985)--(11.947,0.985)--(11.947,8.381)--cycle;
	\node[gp node center,rotate=-270] at (-0.546,4.683) {$D_{KL}\big(p(X_\Theta, X_{\Theta^\prime} \;\vert\; \Theta = \pi) \;\vert\vert\; p(X_\Theta, X_{\Theta^\prime})\big)$};
	\node[gp node center] at (6.725,-0.415) {generation};
	\gpcolor{gp lt color 0}
	\gpsetlinetype{gp lt plot 0}
	\gpsetlinewidth{2.00}
	\draw[gp path] (1.504,6.324)--(1.713,7.767)--(1.922,6.707)--(2.131,5.295)--(2.339,4.588)%
	  --(2.548,4.056)--(2.757,3.801)--(2.966,3.552)--(3.175,2.957)--(3.384,1.296)--(3.593,2.506)%
	  --(3.801,2.506)--(4.010,2.397)--(4.219,2.289)--(4.428,2.077)--(4.637,2.077)--(4.846,1.869)%
	  --(5.055,1.869)--(5.263,1.869)--(5.472,1.767)--(5.681,1.767)--(5.890,1.767)--(6.099,1.767)%
	  --(6.308,1.767)--(6.517,1.767)--(6.726,1.767)--(6.934,1.767)--(7.143,1.666)--(7.352,1.666)%
	  --(7.561,1.666)--(7.770,1.666)--(7.979,1.666)--(8.188,1.666)--(8.396,1.666)--(8.605,1.666)%
	  --(8.814,1.666)--(9.023,1.666)--(9.232,1.666)--(9.441,1.666)--(9.650,1.666)--(9.858,1.666)%
	  --(10.067,1.666)--(10.276,1.666)--(10.485,1.666)--(10.694,1.666)--(10.903,1.666)--(11.112,1.666)%
	  --(11.320,1.666)--(11.529,1.666)--(11.738,1.666)--(11.947,1.666);
	\gpcolor{gp lt color border}
	\gpsetlinetype{gp lt border}
	\gpsetlinewidth{1.00}
	\draw[gp path] (1.504,8.381)--(1.504,0.985)--(11.947,0.985)--(11.947,8.381)--cycle;
	%% coordinates of the plot area
	\gpdefrectangularnode{gp plot 1}{\pgfpoint{1.504cm}{0.985cm}}{\pgfpoint{11.947cm}{8.381cm}}
	\end{tikzpicture}
	%% gnuplot variables
	\caption{\small{We measure how much we can identify the parasite from its messages. As a consequence of maximising damage to the population,
				the parasite blends in. The parasite is introduced at generation $0$.}}
	\label{fig:blendingin}
\end{figure}
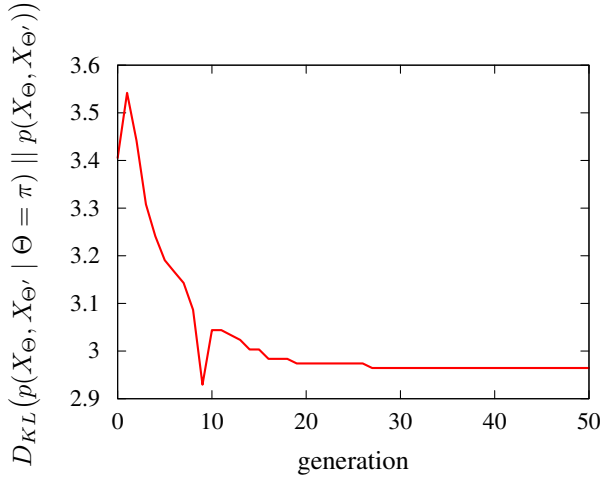

In Fig. \ref{fig:blendingin}, we show the values of Eq. \ref{eq:KLdiv} during the optimisation process, which shows that the divergence diminishes
as the parasite minimises the mutual understanding of the population.

\subsection{Missing environmental aspects}
\label{sec:missing}

However, it is not only a question of choosing common messages: they must not coincide with other agent's codes given the environmental conditions.
In other words, these messages that are popular among the population will be used by the parasite to express different aspects of $\mu$.
Otherwise, if the parasite expresses overlapping aspects, then there might be coincidences with one or more sub-population's adopted conventions.
Consequently, the parasite will capture missing aspects in the population. This can be measured by how much information the code of the parasite
adds about $\mu$ if we look at the average messages. Formally,

\begin{equation}
I\left(\mu \;;\; X_{\Theta^\prime} \;\middle\vert\; X_\Theta, \Theta^\prime = \pi\right)
\label{eq:uniqueaspects}
\end{equation}

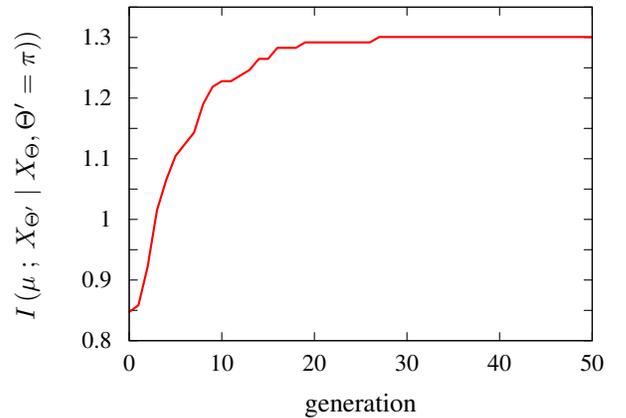
\begin{figure}[ht]
	\centering
	\begin{tikzpicture}[gnuplot, scale = 0.6]
	%% generated with GNUPLOT 4.4p3 (Lua 5.1.4; terminal rev. 97, script rev. 96a)
	%% Tue 17 Mar 2015 04:25:04 GMT
	\gpcolor{gp lt color border}
	\gpsetlinetype{gp lt border}
	\gpsetlinewidth{1.00}
	\draw[gp path] (1.688,0.985)--(1.868,0.985);
	\draw[gp path] (11.947,0.985)--(11.767,0.985);
	\node[gp node right] at (1.504,0.985) {\small{0.8}};
	\draw[gp path] (1.688,1.657)--(1.868,1.657);
	\draw[gp path] (11.947,1.657)--(11.767,1.657);
	%\node[gp node right] at (1.504,1.657) { 0.85};
	\draw[gp path] (1.688,2.330)--(1.868,2.330);
	\draw[gp path] (11.947,2.330)--(11.767,2.330);
	\node[gp node right] at (1.504,2.330) {\small{0.9}};
	\draw[gp path] (1.688,3.002)--(1.868,3.002);
	\draw[gp path] (11.947,3.002)--(11.767,3.002);
	%\node[gp node right] at (1.504,3.002) { 0.95};
	\draw[gp path] (1.688,3.674)--(1.868,3.674);
	\draw[gp path] (11.947,3.674)--(11.767,3.674);
	\node[gp node right] at (1.504,3.674) {\small{1}};
	\draw[gp path] (1.688,4.347)--(1.868,4.347);
	\draw[gp path] (11.947,4.347)--(11.767,4.347);
	%\node[gp node right] at (1.504,4.347) { 1.05};
	\draw[gp path] (1.688,5.019)--(1.868,5.019);
	\draw[gp path] (11.947,5.019)--(11.767,5.019);
	\node[gp node right] at (1.504,5.019) {\small{1.1}};
	\draw[gp path] (1.688,5.692)--(1.868,5.692);
	\draw[gp path] (11.947,5.692)--(11.767,5.692);
	%\node[gp node right] at (1.504,5.692) { 1.15};
	\draw[gp path] (1.688,6.364)--(1.868,6.364);
	\draw[gp path] (11.947,6.364)--(11.767,6.364);
	\node[gp node right] at (1.504,6.364) {\small{1.2}};
	\draw[gp path] (1.688,7.036)--(1.868,7.036);
	\draw[gp path] (11.947,7.036)--(11.767,7.036);
	%\node[gp node right] at (1.504,7.036) { 1.25};
	\draw[gp path] (1.688,7.709)--(1.868,7.709);
	\draw[gp path] (11.947,7.709)--(11.767,7.709);
	\node[gp node right] at (1.504,7.709) {\small{1.3}};
	\draw[gp path] (1.688,8.381)--(1.868,8.381);
	\draw[gp path] (11.947,8.381)--(11.767,8.381);
	%\node[gp node right] at (1.504,8.381) { 1.35};
	\draw[gp path] (1.688,0.985)--(1.688,1.165);
	\draw[gp path] (1.688,8.381)--(1.688,8.201);
	\node[gp node center] at (1.688,0.477) {\small{0}};
	\draw[gp path] (3.740,0.985)--(3.740,1.165);
	\draw[gp path] (3.740,8.381)--(3.740,8.201);
	\node[gp node center] at (3.740,0.477) {\small{10}};
	\draw[gp path] (5.792,0.985)--(5.792,1.165);
	\draw[gp path] (5.792,8.381)--(5.792,8.201);
	\node[gp node center] at (5.792,0.477) {\small{20}};
	\draw[gp path] (7.843,0.985)--(7.843,1.165);
	\draw[gp path] (7.843,8.381)--(7.843,8.201);
	\node[gp node center] at (7.843,0.477) {\small{30}};
	\draw[gp path] (9.895,0.985)--(9.895,1.165);
	\draw[gp path] (9.895,8.381)--(9.895,8.201);
	\node[gp node center] at (9.895,0.477) {\small{40}};
	\draw[gp path] (11.947,0.985)--(11.947,1.165);
	\draw[gp path] (11.947,8.381)--(11.947,8.201);
	\node[gp node center] at (11.947,0.477) {\small{50}};
	\draw[gp path] (1.688,8.381)--(1.688,0.985)--(11.947,0.985)--(11.947,8.381)--cycle;
	\node[gp node center,rotate=-270] at (-0.546,4.683) {$I\left(\mu \;;\; X_{\Theta^\prime} \;\middle\vert\; X_\Theta, \Theta^\prime = \pi)\right)$};
	\node[gp node center] at (6.817,-0.415) {generation};
	\gpcolor{gp lt color 0}
	\gpsetlinetype{gp lt plot 0}
	\gpsetlinewidth{2.00}
	\draw[gp path] (1.688,1.619)--(1.893,1.770)--(2.098,2.629)--(2.304,3.876)--(2.509,4.549)%
	  --(2.714,5.079)--(2.919,5.341)--(3.124,5.598)--(3.329,6.237)--(3.535,6.614)--(3.740,6.739)%
	  --(3.945,6.740)--(4.150,6.864)--(4.355,6.988)--(4.561,7.234)--(4.766,7.234)--(4.971,7.478)%
	  --(5.176,7.478)--(5.381,7.478)--(5.586,7.599)--(5.792,7.599)--(5.997,7.599)--(6.202,7.599)%
	  --(6.407,7.599)--(6.612,7.599)--(6.818,7.599)--(7.023,7.599)--(7.228,7.719)--(7.433,7.719)%
	  --(7.638,7.719)--(7.843,7.719)--(8.049,7.719)--(8.254,7.719)--(8.459,7.719)--(8.664,7.719)%
	  --(8.869,7.719)--(9.074,7.719)--(9.280,7.719)--(9.485,7.719)--(9.690,7.719)--(9.895,7.719)%
	  --(10.100,7.719)--(10.306,7.719)--(10.511,7.719)--(10.716,7.719)--(10.921,7.719)--(11.126,7.719)%
	  --(11.331,7.719)--(11.537,7.719)--(11.742,7.719)--(11.947,7.719);
	\gpcolor{gp lt color border}
	\gpsetlinetype{gp lt border}
	\gpsetlinewidth{1.00}
	\draw[gp path] (1.688,8.381)--(1.688,0.985)--(11.947,0.985)--(11.947,8.381)--cycle;
	%% coordinates of the plot area
	\gpdefrectangularnode{gp plot 1}{\pgfpoint{1.688cm}{0.985cm}}{\pgfpoint{11.947cm}{8.381cm}}
	\end{tikzpicture}
	%% gnuplot variables
	\caption{\small{Amount of information the parasite possesses that the population lacks. The parasite is introduced at generation $0$.}}
	\label{fig:missingaspects}
\end{figure}

The value of Eq. \ref{eq:uniqueaspects} is plotted in Fig. \ref{fig:missingaspects} during the optimisation process. After convergence, we have that 
$I\left(\mu \;;\; X_{\Theta^\prime} \;\middle\vert\; X_\Theta, \Theta^\prime = \pi\right) = 1.30$ bits, while what the parasite acquires, through
its sensors only, is $I\left(\mu \;;\; X_\Theta \;\middle\vert\; \Theta = \pi\right) = 1.32$ bits. That is, almost all the information it captures
is missing in the population. If we consider the perceived information from $\mu$ together with the environmental information provided by the population,
the parasite captures $I\left(\mu \;;\; X_\Theta, X_{\Theta^\prime} \;\middle\vert\; \Theta = \pi\right) = 4$ bits, which is the maximum possible, and
this means that the parasite always correctly predicts the environment.

\subsection{Robustness against parasites}
\label{sec:robustness}

After the parasitic attack, each sub-population has diminished its mutual understanding by different quantities. Although the (former) sub-populations
now share a common agent (the parasite, and thus are not strictly speaking different sub-populations), we maintain the colours used in Fig.
\ref{fig:pardistplot} to identify them. In Table \ref{tab:subpsummary} we show a summary of the outcome of the parasitic attack.

\begin{figure}[ht]
	\centering
	\renewcommand{\arraystretch}{1.5}
		\begin{tabular}{ c  c  c  c  c  c  c }
			Colour & Size & Sizes of types & $I_1$ & $I_2$ & $I_3$ \\
			\hline
			\crule[color1]{1.0cm}{0.5cm} & $47$ & $41, 6$ & $3.93$ & $2.85$ & $3.59$ \\ %I2 = 3.06
			\crule[color2]{1.0cm}{0.5cm} & $35$ & $33, 2$ & $3.93$ & $2.15$ & $3.05$ \\ %I2 = 2.61
			\crule[color3]{1.0cm}{0.5cm} & $16$ & $11, 5$ & $3.93$ & $2.54$ & $3.33$ \\ %I2 = 2.82
			\crule[color4]{1.0cm}{0.5cm} & $8$  & $5 , 3$ & $3.93$ & $2.16$ & $2.98$ \\ %I2 = 2.56
			\crule[color5]{1.0cm}{0.5cm} & $7$  & $5 , 2$ & $3.93$ & $1.83$ & $2.65$ \\ %I2 = 2.35
			\hline
			& $113$ & $95, 18$ & $3.93$ & $2.55$ & $3.50$
		\end{tabular}
		\caption{\small{Summary of the parasitic attack for each sub-population. The colours of each sub-population are the same
						as the ones in Fig. \ref{fig:pardistplot}. $I_1$ is the mutual understanding before the parasitic attack,
						$I_2$ is the mutual understanding after the parasitic attack, and $I_3$ is the mutual understanding after
						the population's response.}}
		\label{tab:subpsummary}
\end{figure}

As Table \ref{tab:subpsummary} shows, in general, larger sub-populations are less damaged by a parasitic intrusion. This phenomenon
is due to the large number of interactions among friendly agents, which diminishes the influence of any single agent by considering the
average of the perceived messages. The exception in the example is the second largest sub-population, which becomes more damaged than the third
largest sub-population. The reason why we see this is that the former sub-population is highly unbalanced, having a small number of agents of
one type. Then, agents of the most numerous type interact only with a small number of agents, and therefore are more vulnerable to malicious agents.

Another way in which a population can defend itself against parasitic attacks is through diversification of their codes. Particularly, agents
can reduce damage by using synonyms to express the same conditions. The presence of synonyms presents an obstacle for the parasite: when trying
to confuse agents by expressing different conditions with a chosen symbol, the meaning of the correspondent synonym is not obfuscated.

We study this by comparing populations with different amounts of code types, while maintaining the same population structure. The setup is the
following: the population is well-mixed (every agent interacts with every other agent), and first we randomly sample a code for every one of the $64$
agents with symbols in the range $[1,16]$. This population has one type of code only, and the used sample has a mutual understanding of $3.5$ bits.
Then, we produce a new population by modifying the code of half of the agents,
such that $p\left(x + 16 \;\middle\vert\; \mu\right) \coloneqq p\left(x \;\middle\vert\; \mu\right)$ and then we set $p\left(x \;\middle\vert\; \mu\right) \coloneqq 0$ 
(here, $x + 16$ is a synonym of $x$). In this way, the mutual understanding remains the same for the modified population, which now has two types.
Each new type is introduced in a similar fashion, always mapping the original code to a set of ($16$) unused symbols.

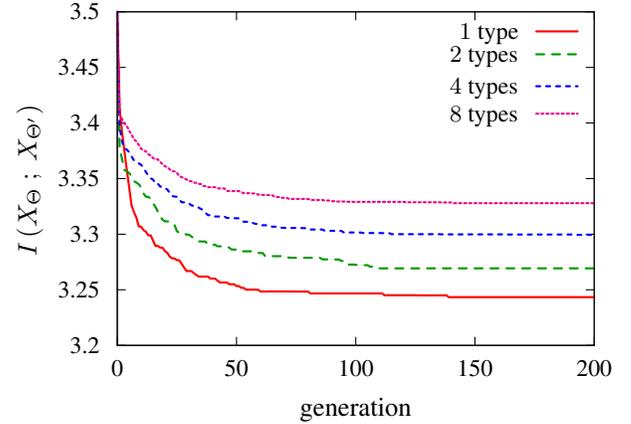
\begin{figure}[ht]
	\centering
	\begin{tikzpicture}[gnuplot, scale = 0.6]
	%% generated with GNUPLOT 4.4p3 (Lua 5.1.4; terminal rev. 97, script rev. 96a)
	%% Tue 17 Mar 2015 04:32:24 GMT
	\gpcolor{gp lt color border}
	\gpsetlinetype{gp lt border}
	\gpsetlinewidth{1.00}
	\draw[gp path] (1.380,0.985)--(1.560,0.985);
	\draw[gp path] (11.947,0.985)--(11.767,0.985);
	\node[gp node right] at (1.196,0.985) {\small{3.2}};
	\draw[gp path] (1.380,2.218)--(1.560,2.218);
	\draw[gp path] (11.947,2.218)--(11.767,2.218);
	\node[gp node right] at (1.196,2.218) {\small{3.25}};
	\draw[gp path] (1.380,3.450)--(1.560,3.450);
	\draw[gp path] (11.947,3.450)--(11.767,3.450);
	\node[gp node right] at (1.196,3.450) {\small{3.3}};
	\draw[gp path] (1.380,4.683)--(1.560,4.683);
	\draw[gp path] (11.947,4.683)--(11.767,4.683);
	\node[gp node right] at (1.196,4.683) { \small{3.35}};
	\draw[gp path] (1.380,5.916)--(1.560,5.916);
	\draw[gp path] (11.947,5.916)--(11.767,5.916);
	\node[gp node right] at (1.196,5.916) { \small{3.4}};
	\draw[gp path] (1.380,7.148)--(1.560,7.148);
	\draw[gp path] (11.947,7.148)--(11.767,7.148);
	\node[gp node right] at (1.196,7.148) { \small{3.45}};
	\draw[gp path] (1.380,8.381)--(1.560,8.381);
	\draw[gp path] (11.947,8.381)--(11.767,8.381);
	\node[gp node right] at (1.196,8.381) { \small{3.5}};
	\draw[gp path] (1.380,0.985)--(1.380,1.165);
	\draw[gp path] (1.380,8.381)--(1.380,8.201);
	\node[gp node center] at (1.380,0.477) {\small{0}};
	\draw[gp path] (4.022,0.985)--(4.022,1.165);
	\draw[gp path] (4.022,8.381)--(4.022,8.201);
	\node[gp node center] at (4.022,0.477) {\small{50}};
	\draw[gp path] (6.664,0.985)--(6.664,1.165);
	\draw[gp path] (6.664,8.381)--(6.664,8.201);
	\node[gp node center] at (6.664,0.477) {\small{100}};
	\draw[gp path] (9.305,0.985)--(9.305,1.165);
	\draw[gp path] (9.305,8.381)--(9.305,8.201);
	\node[gp node center] at (9.305,0.477) {\small{150}};
	\draw[gp path] (11.947,0.985)--(11.947,1.165);
	\draw[gp path] (11.947,8.381)--(11.947,8.201);
	\node[gp node center] at (11.947,0.477) {\small{200}};
	\draw[gp path] (1.380,8.381)--(1.380,0.985)--(11.947,0.985)--(11.947,8.381)--cycle;
	\node[gp node center,rotate=-270] at (-0.546,4.683) {$I\left(X_\Theta \;;\; X_{\Theta^\prime}\right)$};
	\node[gp node center] at (6.663,-0.415) {generation};
	\node[gp node right] at (10.479,7.947) {\small{$1$ type}};
	\gpcolor{gp lt color 0}
	\gpsetlinetype{gp lt plot 0}
	\gpsetlinewidth{2.00}
	\draw[gp path] (10.663,7.947)--(11.579,7.947);
	\draw[gp path] (1.380,8.381)--(1.433,6.250)--(1.486,5.710)--(1.539,5.283)--(1.591,4.876)%
	  --(1.644,4.486)--(1.697,4.086)--(1.750,3.939)--(1.803,3.818)--(1.856,3.621)--(1.908,3.618)%
	  --(1.961,3.541)--(2.014,3.504)--(2.067,3.425)--(2.120,3.425)--(2.173,3.308)--(2.225,3.190)%
	  --(2.278,3.190)--(2.331,3.152)--(2.384,3.152)--(2.437,3.073)--(2.490,3.034)--(2.542,2.954)%
	  --(2.595,2.913)--(2.648,2.913)--(2.701,2.873)--(2.754,2.872)--(2.807,2.793)--(2.859,2.711)%
	  --(2.912,2.632)--(2.965,2.632)--(3.018,2.632)--(3.071,2.591)--(3.124,2.551)--(3.176,2.512)%
	  --(3.229,2.512)--(3.282,2.510)--(3.335,2.510)--(3.388,2.510)--(3.441,2.466)--(3.493,2.466)%
	  --(3.546,2.466)--(3.599,2.421)--(3.652,2.380)--(3.705,2.380)--(3.758,2.380)--(3.810,2.380)%
	  --(3.863,2.340)--(3.916,2.340)--(3.969,2.340)--(4.022,2.300)--(4.075,2.300)--(4.127,2.260)%
	  --(4.180,2.260)--(4.233,2.220)--(4.286,2.220)--(4.339,2.220)--(4.392,2.220)--(4.444,2.220)%
	  --(4.497,2.220)--(4.550,2.181)--(4.603,2.181)--(4.656,2.180)--(4.709,2.180)--(4.761,2.180)%
	  --(4.814,2.180)--(4.867,2.180)--(4.920,2.180)--(4.973,2.180)--(5.026,2.180)--(5.078,2.180)%
	  --(5.131,2.180)--(5.184,2.180)--(5.237,2.178)--(5.290,2.178)--(5.343,2.178)--(5.395,2.178)%
	  --(5.448,2.177)--(5.501,2.177)--(5.554,2.177)--(5.607,2.177)--(5.660,2.137)--(5.712,2.137)%
	  --(5.765,2.137)--(5.818,2.137)--(5.871,2.137)--(5.924,2.137)--(5.977,2.137)--(6.029,2.137)%
	  --(6.082,2.137)--(6.135,2.137)--(6.188,2.137)--(6.241,2.137)--(6.294,2.137)--(6.346,2.137)%
	  --(6.399,2.137)--(6.452,2.137)--(6.505,2.137)--(6.558,2.137)--(6.611,2.137)--(6.664,2.137)%
	  --(6.716,2.137)--(6.769,2.137)--(6.822,2.137)--(6.875,2.137)--(6.928,2.137)--(6.981,2.137)%
	  --(7.033,2.137)--(7.086,2.137)--(7.139,2.137)--(7.192,2.137)--(7.245,2.137)--(7.298,2.097)%
	  --(7.350,2.097)--(7.403,2.097)--(7.456,2.097)--(7.509,2.097)--(7.562,2.097)--(7.615,2.097)%
	  --(7.667,2.097)--(7.720,2.097)--(7.773,2.097)--(7.826,2.097)--(7.879,2.097)--(7.932,2.097)%
	  --(7.984,2.097)--(8.037,2.094)--(8.090,2.094)--(8.143,2.094)--(8.196,2.094)--(8.249,2.094)%
	  --(8.301,2.094)--(8.354,2.094)--(8.407,2.094)--(8.460,2.094)--(8.513,2.094)--(8.566,2.094)%
	  --(8.618,2.094)--(8.671,2.094)--(8.724,2.054)--(8.777,2.054)--(8.830,2.054)--(8.883,2.054)%
	  --(8.935,2.054)--(8.988,2.054)--(9.041,2.054)--(9.094,2.054)--(9.147,2.054)--(9.200,2.054)%
	  --(9.252,2.054)--(9.305,2.054)--(9.358,2.054)--(9.411,2.054)--(9.464,2.054)--(9.517,2.054)%
	  --(9.569,2.054)--(9.622,2.054)--(9.675,2.054)--(9.728,2.054)--(9.781,2.054)--(9.834,2.054)%
	  --(9.886,2.054)--(9.939,2.054)--(9.992,2.054)--(10.045,2.054)--(10.098,2.054)--(10.151,2.054)%
	  --(10.203,2.054)--(10.256,2.054)--(10.309,2.054)--(10.362,2.054)--(10.415,2.054)--(10.468,2.054)%
	  --(10.520,2.054)--(10.573,2.054)--(10.626,2.054)--(10.679,2.054)--(10.732,2.054)--(10.785,2.054)%
	  --(10.837,2.054)--(10.890,2.054)--(10.943,2.054)--(10.996,2.054)--(11.049,2.054)--(11.102,2.054)%
	  --(11.154,2.054)--(11.207,2.054)--(11.260,2.054)--(11.313,2.054)--(11.366,2.054)--(11.419,2.054)%
	  --(11.471,2.054)--(11.524,2.054)--(11.577,2.054)--(11.630,2.054)--(11.683,2.054)--(11.736,2.054)%
	  --(11.788,2.054)--(11.841,2.054)--(11.894,2.054)--(11.947,2.054);
	\gpcolor{gp lt color border}
	\node[gp node right] at (10.479,7.439) {\small{$2$ types}};
	\gpcolor{gp lt color 1}
	\gpsetlinetype{gp lt plot 1}
	\draw[gp path] (10.663,7.439)--(11.579,7.439);
	\draw[gp path] (1.380,8.381)--(1.433,5.348)--(1.486,5.115)--(1.539,4.872)--(1.591,4.843)%
	  --(1.644,4.806)--(1.697,4.733)--(1.750,4.626)--(1.803,4.594)--(1.856,4.558)--(1.908,4.489)%
	  --(1.961,4.353)--(2.014,4.280)--(2.067,4.279)--(2.120,4.239)--(2.173,4.132)--(2.225,4.058)%
	  --(2.278,3.953)--(2.331,3.845)--(2.384,3.771)--(2.437,3.734)--(2.490,3.734)--(2.542,3.732)%
	  --(2.595,3.698)--(2.648,3.660)--(2.701,3.516)--(2.754,3.481)--(2.807,3.476)--(2.859,3.476)%
	  --(2.912,3.440)--(2.965,3.440)--(3.018,3.403)--(3.071,3.363)--(3.124,3.326)--(3.176,3.291)%
	  --(3.229,3.290)--(3.282,3.288)--(3.335,3.288)--(3.388,3.217)--(3.441,3.217)--(3.493,3.217)%
	  --(3.546,3.217)--(3.599,3.182)--(3.652,3.182)--(3.705,3.182)--(3.758,3.182)--(3.810,3.182)%
	  --(3.863,3.147)--(3.916,3.110)--(3.969,3.110)--(4.022,3.110)--(4.075,3.071)--(4.127,3.071)%
	  --(4.180,3.071)--(4.233,3.071)--(4.286,3.071)--(4.339,3.070)--(4.392,3.070)--(4.444,3.070)%
	  --(4.497,3.035)--(4.550,3.033)--(4.603,3.033)--(4.656,2.963)--(4.709,2.963)--(4.761,2.963)%
	  --(4.814,2.963)--(4.867,2.963)--(4.920,2.963)--(4.973,2.963)--(5.026,2.963)--(5.078,2.963)%
	  --(5.131,2.963)--(5.184,2.927)--(5.237,2.927)--(5.290,2.927)--(5.343,2.927)--(5.395,2.927)%
	  --(5.448,2.927)--(5.501,2.927)--(5.554,2.927)--(5.607,2.927)--(5.660,2.927)--(5.712,2.927)%
	  --(5.765,2.927)--(5.818,2.927)--(5.871,2.927)--(5.924,2.892)--(5.977,2.892)--(6.029,2.892)%
	  --(6.082,2.892)--(6.135,2.889)--(6.188,2.889)--(6.241,2.889)--(6.294,2.889)--(6.346,2.855)%
	  --(6.399,2.855)--(6.452,2.855)--(6.505,2.777)--(6.558,2.777)--(6.611,2.777)--(6.664,2.771)%
	  --(6.716,2.763)--(6.769,2.763)--(6.822,2.763)--(6.875,2.763)--(6.928,2.763)--(6.981,2.728)%
	  --(7.033,2.728)--(7.086,2.728)--(7.139,2.693)--(7.192,2.693)--(7.245,2.693)--(7.298,2.693)%
	  --(7.350,2.693)--(7.403,2.693)--(7.456,2.693)--(7.509,2.693)--(7.562,2.693)--(7.615,2.693)%
	  --(7.667,2.693)--(7.720,2.693)--(7.773,2.693)--(7.826,2.693)--(7.879,2.693)--(7.932,2.693)%
	  --(7.984,2.693)--(8.037,2.693)--(8.090,2.693)--(8.143,2.693)--(8.196,2.693)--(8.249,2.693)%
	  --(8.301,2.693)--(8.354,2.693)--(8.407,2.693)--(8.460,2.693)--(8.513,2.693)--(8.566,2.693)%
	  --(8.618,2.693)--(8.671,2.693)--(8.724,2.693)--(8.777,2.693)--(8.830,2.693)--(8.883,2.693)%
	  --(8.935,2.693)--(8.988,2.693)--(9.041,2.693)--(9.094,2.693)--(9.147,2.693)--(9.200,2.693)%
	  --(9.252,2.693)--(9.305,2.693)--(9.358,2.693)--(9.411,2.693)--(9.464,2.693)--(9.517,2.693)%
	  --(9.569,2.693)--(9.622,2.693)--(9.675,2.693)--(9.728,2.693)--(9.781,2.693)--(9.834,2.693)%
	  --(9.886,2.693)--(9.939,2.693)--(9.992,2.693)--(10.045,2.693)--(10.098,2.693)--(10.151,2.693)%
	  --(10.203,2.693)--(10.256,2.693)--(10.309,2.693)--(10.362,2.693)--(10.415,2.693)--(10.468,2.693)%
	  --(10.520,2.693)--(10.573,2.693)--(10.626,2.693)--(10.679,2.693)--(10.732,2.693)--(10.785,2.693)%
	  --(10.837,2.693)--(10.890,2.693)--(10.943,2.693)--(10.996,2.693)--(11.049,2.693)--(11.102,2.693)%
	  --(11.154,2.693)--(11.207,2.693)--(11.260,2.693)--(11.313,2.693)--(11.366,2.693)--(11.419,2.693)%
	  --(11.471,2.693)--(11.524,2.693)--(11.577,2.693)--(11.630,2.693)--(11.683,2.693)--(11.736,2.693)%
	  --(11.788,2.693)--(11.841,2.693)--(11.894,2.693)--(11.947,2.693);
	\gpcolor{gp lt color border}
	\node[gp node right] at (10.479,6.731) {\small{$4$ types}};
	\gpcolor{gp lt color 2}
	\gpsetlinetype{gp lt plot 2}
	\draw[gp path] (10.663,6.731)--(11.579,6.731);
	\draw[gp path] (1.380,8.381)--(1.433,5.817)--(1.486,5.509)--(1.539,5.353)--(1.591,5.326)%
	  --(1.644,5.267)--(1.697,5.175)--(1.750,5.086)--(1.803,5.055)--(1.856,5.021)--(1.908,4.988)%
	  --(1.961,4.925)--(2.014,4.829)--(2.067,4.753)--(2.120,4.689)--(2.173,4.661)--(2.225,4.573)%
	  --(2.278,4.573)--(2.331,4.504)--(2.384,4.504)--(2.437,4.470)--(2.490,4.439)--(2.542,4.365)%
	  --(2.595,4.296)--(2.648,4.285)--(2.701,4.279)--(2.754,4.220)--(2.807,4.157)--(2.859,4.157)%
	  --(2.912,4.138)--(2.965,4.138)--(3.018,4.092)--(3.071,4.077)--(3.124,4.047)--(3.176,4.044)%
	  --(3.229,4.013)--(3.282,3.984)--(3.335,3.925)--(3.388,3.872)--(3.441,3.872)--(3.493,3.842)%
	  --(3.546,3.842)--(3.599,3.842)--(3.652,3.842)--(3.705,3.840)--(3.758,3.839)--(3.810,3.809)%
	  --(3.863,3.809)--(3.916,3.804)--(3.969,3.804)--(4.022,3.804)--(4.075,3.802)--(4.127,3.772)%
	  --(4.180,3.725)--(4.233,3.725)--(4.286,3.725)--(4.339,3.697)--(4.392,3.697)--(4.444,3.678)%
	  --(4.497,3.678)--(4.550,3.648)--(4.603,3.648)--(4.656,3.648)--(4.709,3.648)--(4.761,3.648)%
	  --(4.814,3.616)--(4.867,3.616)--(4.920,3.616)--(4.973,3.616)--(5.026,3.591)--(5.078,3.591)%
	  --(5.131,3.591)--(5.184,3.590)--(5.237,3.588)--(5.290,3.588)--(5.343,3.588)--(5.395,3.588)%
	  --(5.448,3.588)--(5.501,3.584)--(5.554,3.584)--(5.607,3.554)--(5.660,3.554)--(5.712,3.554)%
	  --(5.765,3.554)--(5.818,3.554)--(5.871,3.554)--(5.924,3.554)--(5.977,3.527)--(6.029,3.527)%
	  --(6.082,3.525)--(6.135,3.525)--(6.188,3.525)--(6.241,3.523)--(6.294,3.523)--(6.346,3.523)%
	  --(6.399,3.488)--(6.452,3.488)--(6.505,3.488)--(6.558,3.488)--(6.611,3.488)--(6.664,3.488)%
	  --(6.716,3.488)--(6.769,3.488)--(6.822,3.476)--(6.875,3.476)--(6.928,3.476)--(6.981,3.476)%
	  --(7.033,3.476)--(7.086,3.476)--(7.139,3.476)--(7.192,3.476)--(7.245,3.474)--(7.298,3.474)%
	  --(7.350,3.474)--(7.403,3.474)--(7.456,3.450)--(7.509,3.450)--(7.562,3.450)--(7.615,3.450)%
	  --(7.667,3.450)--(7.720,3.450)--(7.773,3.450)--(7.826,3.450)--(7.879,3.450)--(7.932,3.450)%
	  --(7.984,3.450)--(8.037,3.450)--(8.090,3.450)--(8.143,3.450)--(8.196,3.450)--(8.249,3.450)%
	  --(8.301,3.450)--(8.354,3.450)--(8.407,3.450)--(8.460,3.450)--(8.513,3.450)--(8.566,3.446)%
	  --(8.618,3.446)--(8.671,3.446)--(8.724,3.446)--(8.777,3.446)--(8.830,3.446)--(8.883,3.446)%
	  --(8.935,3.446)--(8.988,3.446)--(9.041,3.446)--(9.094,3.446)--(9.147,3.446)--(9.200,3.446)%
	  --(9.252,3.446)--(9.305,3.446)--(9.358,3.446)--(9.411,3.446)--(9.464,3.446)--(9.517,3.446)%
	  --(9.569,3.446)--(9.622,3.446)--(9.675,3.446)--(9.728,3.446)--(9.781,3.446)--(9.834,3.446)%
	  --(9.886,3.446)--(9.939,3.446)--(9.992,3.446)--(10.045,3.443)--(10.098,3.443)--(10.151,3.443)%
	  --(10.203,3.443)--(10.256,3.443)--(10.309,3.443)--(10.362,3.443)--(10.415,3.443)--(10.468,3.443)%
	  --(10.520,3.443)--(10.573,3.443)--(10.626,3.443)--(10.679,3.443)--(10.732,3.443)--(10.785,3.443)%
	  --(10.837,3.443)--(10.890,3.443)--(10.943,3.443)--(10.996,3.443)--(11.049,3.443)--(11.102,3.443)%
	  --(11.154,3.443)--(11.207,3.443)--(11.260,3.443)--(11.313,3.443)--(11.366,3.443)--(11.419,3.443)%
	  --(11.471,3.443)--(11.524,3.436)--(11.577,3.436)--(11.630,3.436)--(11.683,3.436)--(11.736,3.436)%
	  --(11.788,3.436)--(11.841,3.436)--(11.894,3.436)--(11.947,3.436);
	\gpcolor{gp lt color border}
	\node[gp node right] at (10.479,6.123) {\small{$8$ types}};
	\gpcolor{gp lt color 3}
	\gpsetlinetype{gp lt plot 3}
	\draw[gp path] (10.663,6.123)--(11.579,6.123);
	\draw[gp path] (1.380,8.381)--(1.433,6.106)--(1.486,6.002)--(1.539,5.902)--(1.591,5.849)%
	  --(1.644,5.744)--(1.697,5.660)--(1.750,5.553)--(1.803,5.508)--(1.856,5.429)--(1.908,5.346)%
	  --(1.961,5.311)--(2.014,5.287)--(2.067,5.227)--(2.120,5.188)--(2.173,5.138)--(2.225,5.110)%
	  --(2.278,5.110)--(2.331,5.054)--(2.384,4.998)--(2.437,4.957)--(2.490,4.934)--(2.542,4.887)%
	  --(2.595,4.846)--(2.648,4.843)--(2.701,4.813)--(2.754,4.738)--(2.807,4.715)--(2.859,4.690)%
	  --(2.912,4.666)--(2.965,4.627)--(3.018,4.627)--(3.071,4.591)--(3.124,4.591)--(3.176,4.565)%
	  --(3.229,4.565)--(3.282,4.517)--(3.335,4.517)--(3.388,4.492)--(3.441,4.492)--(3.493,4.492)%
	  --(3.546,4.492)--(3.599,4.468)--(3.652,4.465)--(3.705,4.465)--(3.758,4.464)--(3.810,4.409)%
	  --(3.863,4.406)--(3.916,4.406)--(3.969,4.406)--(4.022,4.406)--(4.075,4.406)--(4.127,4.381)%
	  --(4.180,4.360)--(4.233,4.360)--(4.286,4.360)--(4.339,4.360)--(4.392,4.353)--(4.444,4.353)%
	  --(4.497,4.329)--(4.550,4.329)--(4.603,4.320)--(4.656,4.320)--(4.709,4.320)--(4.761,4.313)%
	  --(4.814,4.313)--(4.867,4.281)--(4.920,4.274)--(4.973,4.274)--(5.026,4.257)--(5.078,4.257)%
	  --(5.131,4.257)--(5.184,4.232)--(5.237,4.232)--(5.290,4.232)--(5.343,4.232)--(5.395,4.232)%
	  --(5.448,4.232)--(5.501,4.232)--(5.554,4.232)--(5.607,4.232)--(5.660,4.208)--(5.712,4.208)%
	  --(5.765,4.208)--(5.818,4.208)--(5.871,4.208)--(5.924,4.208)--(5.977,4.196)--(6.029,4.196)%
	  --(6.082,4.196)--(6.135,4.182)--(6.188,4.182)--(6.241,4.175)--(6.294,4.175)--(6.346,4.175)%
	  --(6.399,4.175)--(6.452,4.175)--(6.505,4.167)--(6.558,4.167)--(6.611,4.167)--(6.664,4.167)%
	  --(6.716,4.167)--(6.769,4.167)--(6.822,4.167)--(6.875,4.167)--(6.928,4.167)--(6.981,4.167)%
	  --(7.033,4.167)--(7.086,4.167)--(7.139,4.167)--(7.192,4.167)--(7.245,4.167)--(7.298,4.167)%
	  --(7.350,4.167)--(7.403,4.167)--(7.456,4.167)--(7.509,4.167)--(7.562,4.167)--(7.615,4.167)%
	  --(7.667,4.167)--(7.720,4.167)--(7.773,4.158)--(7.826,4.158)--(7.879,4.158)--(7.932,4.158)%
	  --(7.984,4.158)--(8.037,4.158)--(8.090,4.158)--(8.143,4.158)--(8.196,4.158)--(8.249,4.158)%
	  --(8.301,4.158)--(8.354,4.158)--(8.407,4.158)--(8.460,4.158)--(8.513,4.158)--(8.566,4.154)%
	  --(8.618,4.154)--(8.671,4.154)--(8.724,4.154)--(8.777,4.154)--(8.830,4.138)--(8.883,4.138)%
	  --(8.935,4.138)--(8.988,4.138)--(9.041,4.138)--(9.094,4.138)--(9.147,4.138)--(9.200,4.138)%
	  --(9.252,4.138)--(9.305,4.138)--(9.358,4.138)--(9.411,4.138)--(9.464,4.138)--(9.517,4.138)%
	  --(9.569,4.138)--(9.622,4.138)--(9.675,4.138)--(9.728,4.138)--(9.781,4.138)--(9.834,4.138)%
	  --(9.886,4.138)--(9.939,4.138)--(9.992,4.138)--(10.045,4.138)--(10.098,4.138)--(10.151,4.138)%
	  --(10.203,4.138)--(10.256,4.138)--(10.309,4.138)--(10.362,4.138)--(10.415,4.138)--(10.468,4.138)%
	  --(10.520,4.138)--(10.573,4.138)--(10.626,4.138)--(10.679,4.138)--(10.732,4.138)--(10.785,4.138)%
	  --(10.837,4.138)--(10.890,4.138)--(10.943,4.138)--(10.996,4.138)--(11.049,4.138)--(11.102,4.138)%
	  --(11.154,4.138)--(11.207,4.138)--(11.260,4.138)--(11.313,4.138)--(11.366,4.138)--(11.419,4.138)%
	  --(11.471,4.138)--(11.524,4.138)--(11.577,4.138)--(11.630,4.138)--(11.683,4.138)--(11.736,4.138)%
	  --(11.788,4.138)--(11.841,4.138)--(11.894,4.138)--(11.947,4.138);
	\gpcolor{gp lt color border}
	\gpsetlinetype{gp lt border}
	\gpsetlinewidth{1.00}
	\draw[gp path] (1.380,8.381)--(1.380,0.985)--(11.947,0.985)--(11.947,8.381)--cycle;
	%% coordinates of the plot area
	\gpdefrectangularnode{gp plot 1}{\pgfpoint{1.380cm}{0.985cm}}{\pgfpoint{11.947cm}{8.381cm}}
	\end{tikzpicture}
	\caption{\small{Mutual understanding of populations with varying amounts of types of codes during optimisation. Populations with more types of
				codes are more resistant to parasitic attacks. In all cases, the parasite was introduced at generation $0$.}}
	\label{fig:synonyms}
\end{figure}

We perform the minimisation of the mutual understanding on each population by introducing a parasite until convergence. We show in Fig.
\ref{fig:synonyms} the values of $I(X_\Theta \;;\; X_{\Theta^\prime})$ during the optimisation process for each population. Our results confirm
our expectations: more diverse populations are more resistant to parasitic attacks.

%Let us recall that when the mutual understanding is (locally) optimal, then agents within a sub-population
%necessarily capture the same aspects of the environment, and the mapping $p\left(X_\Theta \;\middle\vert\; X_{\Theta^\prime}\right)$ is deterministic.
%Therefore, if $p(x, x^\prime) > 0$, then $x$ is a synonym of $x^\prime$ within the sub-population (they both refer to the same environmental condition).

\subsection{Code diversification}
\label{sec:armsrace}

Now we let the population respond to the parasitic attack. If we let the rest of the agents respond to the parasite by freely changing the
structure of the population, then our simulations shows that the parasite becomes isolated from the population, which is the expected outcome.
However, we consider here a scenario where the structure of the population is maintained, and agents can only respond to the attack by updating
their codes. In the same way as we did with the parasite, we allow the agents to choose from a larger set of messages (we consider $32$ symbols
to give the option to agents of changing completely their encoding of $\mu$).

\begin{figure}[htp]
	\centering
	\begin{minipage}[t]{.23\textwidth}
		\includegraphics[scale=0.13]{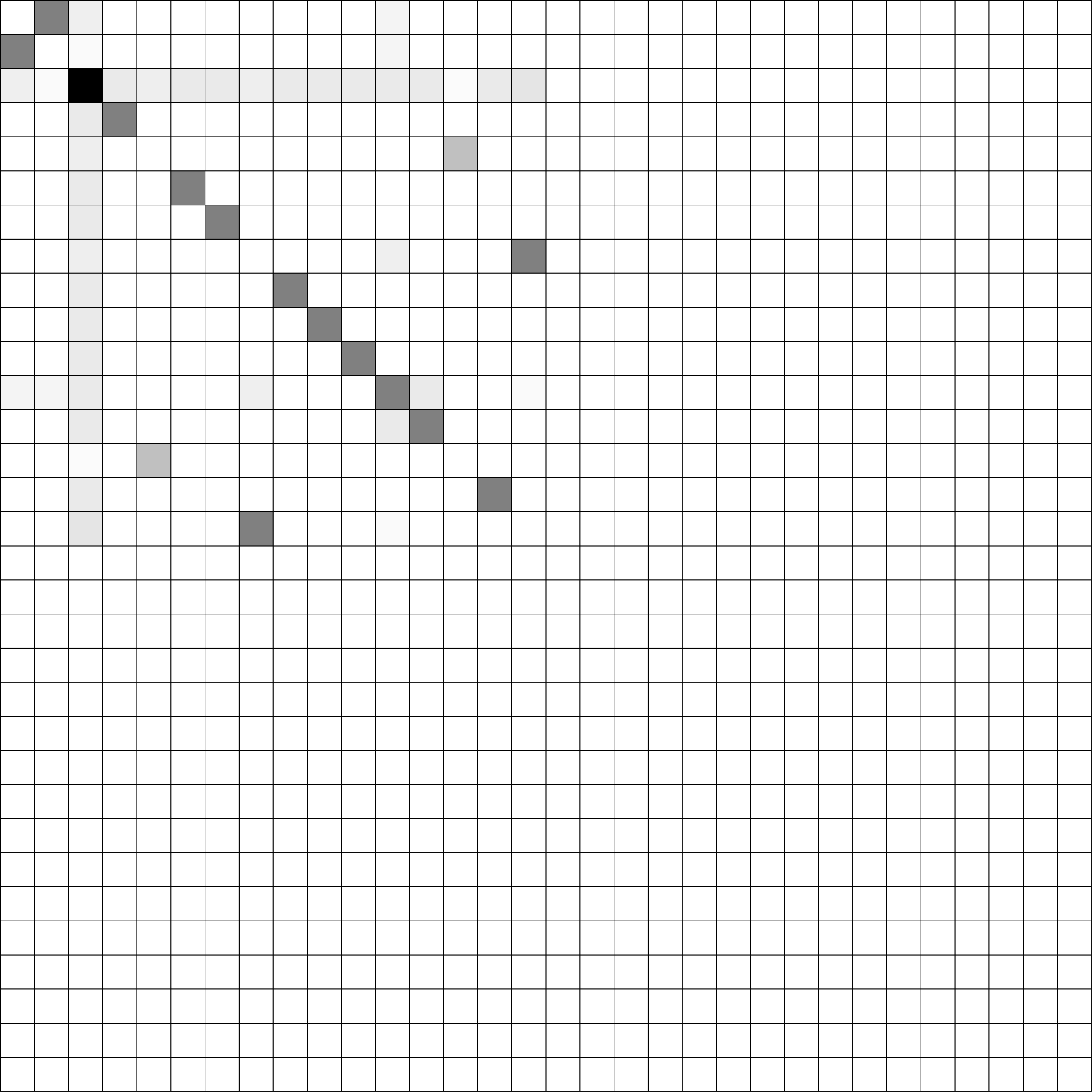}
		\subcaption{\small{after the parasitic attack.}}
	\end{minipage}\hfill
	\begin{minipage}[t]{.23\textwidth}
		\includegraphics[scale=0.13]{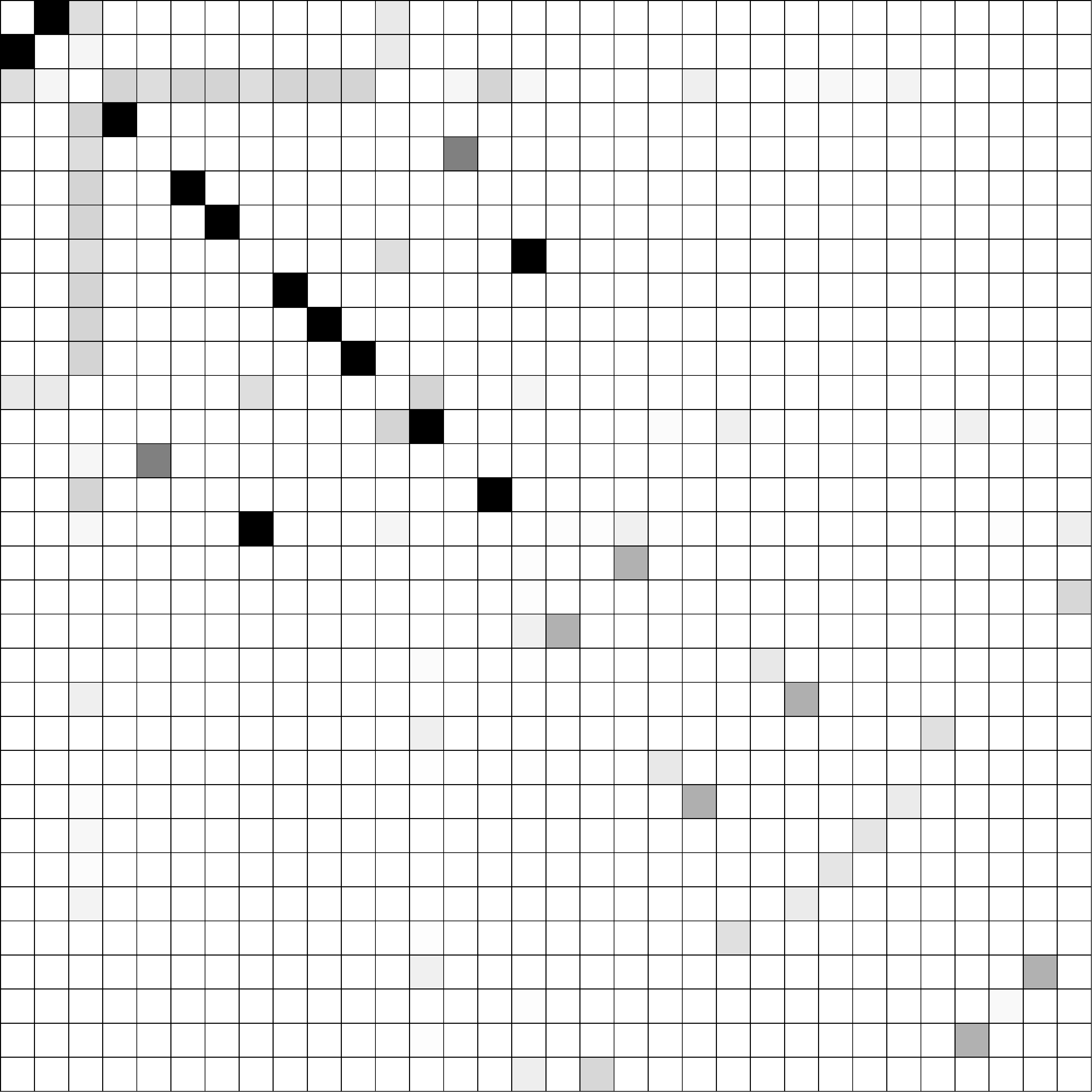}
		\subcaption{\small{after the population's response.}}
	\end{minipage}
	\caption{\small{Joint probability of messages $p(X_\Theta, X_{\Theta^\prime})$ (a) after the parasitic attack and (b) after
					the population's response.}}
	\label{fig:jointprobs}
\end{figure}

After convergence, the population's mutual understanding recovered to a value of $I\left(X_\Theta \;;\; X_{\Theta^\prime}\right) = 3.50$ bits (see
Table \ref{tab:subpsummary} for a summary). In response to the parasitic attack, the agents of the population replaced, mostly, the symbols utilised
by the parasite with unused ones. In Fig. \ref{fig:jointprobs}, we can see how the joint probability $p(X_\Theta, X_{\Theta^\prime})$ changed after
the population's response. Here, the symbols present in the parasite's code ($x_3$, $x_{12}$, $x_{13}$ and $x_{16}$) are mostly removed from the
population's codes.% (these messages still occur in some agents).

\begin{figure}[ht]
	\centering
	\includegraphics[scale=0.5]{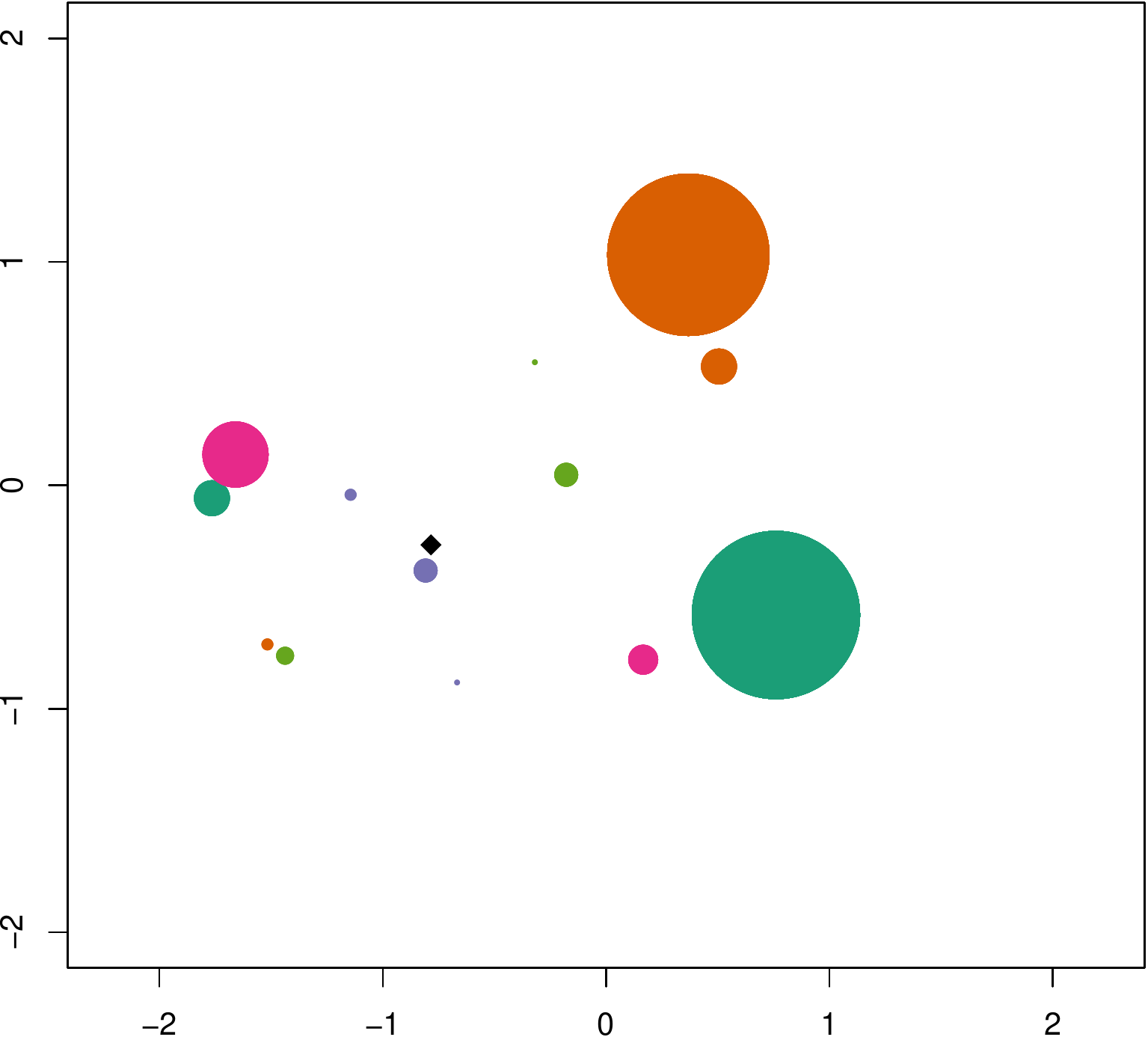}
	\caption{\small{2-dimensional plot of the distance between the codes after the population's response to the parasite attack. Each point
			represents a particular code, and its size is relative to the number of agents adopting that particular code. The colour of the
			points denotes the sub-population to which the codes belong. The black diamond represents the parasite's code.}}
	\label{fig:pardistrespplot}
\end{figure}

Three important features follow from the population's response: first, new code profiles are created in the population. For instance, the
orange, purple and green sub-populations shown in Fig. \ref{fig:pardistplot} now consist of three types of codes (see Fig. \ref{fig:pardistrespplot}.
Nevertheless, the bipartite property is kept, but, instead, synonyms were adopted by one type in these sub-populations. This is due to the large
amount of symbols available that, in the case they are not in use within the agent's type, are detached from any meaning and thus would not create
confusion.

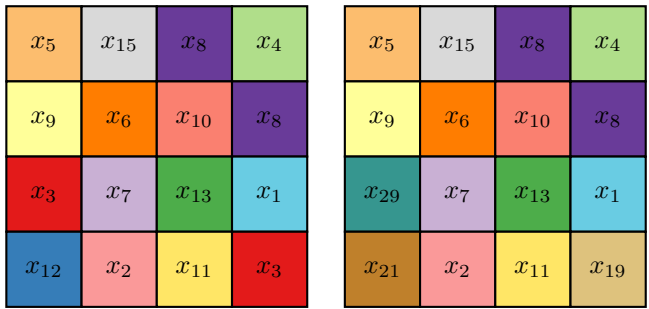
\begin{figure}[ht]
	\centering
	\begin{tikzpicture}
		\def\bc{3.5} % space between codes
		\def\as{0} % added space for each code
		% code number
		\def\code{0}
		\filldraw[draw=black,thick,fill={rgb,1:red,0.99;green,0.74;blue,0.43}] (0+\code+\code*\bc,-0) rectangle (1.0+\code+\code*\bc,-1.0);
		\filldraw[draw=black,thick,fill={rgb,1:red,0.85;green,0.85;blue,0.85}] (1.0+\code+\code*\bc,-0) rectangle (2.0+\code+\code*\bc,-1.0);
		\filldraw[draw=black,thick,fill={rgb,1:red,0.41;green,0.23;blue,0.60}] (2.0+\code+\code*\bc,-0) rectangle (3.0+\code+\code*\bc,-1.0);
		\filldraw[draw=black,thick,fill={rgb,1:red,0.69;green,0.87;blue,0.54}] (3.0+\code+\code*\bc,-0) rectangle (4.0+\code+\code*\bc,-1.0);
		\filldraw[draw=black,thick,fill={rgb,1:red,1;green,1;blue,0.60}] (0+\code+\code*\bc,-1.0) rectangle (1.0+\code+\code*\bc,-2.0);
		\filldraw[draw=black,thick,fill={rgb,1:red,1;green,0.49;blue,0}] (1.0+\code+\code*\bc,-1.0) rectangle (2.0+\code+\code*\bc,-2.0);
		\filldraw[draw=black,thick,fill={rgb,1:red,0.98;green,0.50;blue,0.44}] (2.0+\code+\code*\bc,-1.0) rectangle (3.0+\code+\code*\bc,-2.0);
		\filldraw[draw=black,thick,fill={rgb,1:red,0.41;green,0.23;blue,0.60}] (3.0+\code+\code*\bc,-1.0) rectangle (4.0+\code+\code*\bc,-2.0);
		\filldraw[draw=black,thick,fill={rgb,1:red,0.89;green,0.1;blue,0.1}] (0+\code+\code*\bc,-2.0) rectangle (1.0+\code+\code*\bc,-3.0);
		\filldraw[draw=black,thick,fill={rgb,1:red,0.79;green,0.69;blue,0.83}] (1.0+\code+\code*\bc,-2.0) rectangle (2.0+\code+\code*\bc,-3.0);
		\filldraw[draw=black,thick,fill={rgb,1:red,0.30;green,0.68;blue,0.29}] (2.0+\code+\code*\bc,-2.0) rectangle (3.0+\code+\code*\bc,-3.0);
		\filldraw[draw=black,thick,fill={rgb,1:red,0.41;green,0.80;blue,0.89}] (3.0+\code+\code*\bc,-2.0) rectangle (4.0+\code+\code*\bc,-3.0);
		\filldraw[draw=black,thick,fill={rgb,1:red,0.21;green,0.49;blue,0.72}] (0+\code+\code*\bc,-3.0) rectangle (1.0+\code+\code*\bc,-4.0);
		\filldraw[draw=black,thick,fill={rgb,1:red,0.98;green,0.60;blue,0.60}] (1.0+\code+\code*\bc,-3.0) rectangle (2.0+\code+\code*\bc,-4.0);
		\filldraw[draw=black,thick,fill={rgb,1:red,1;green,0.90;blue,0.40}] (2.0+\code+\code*\bc,-3.0) rectangle (3.0+\code+\code*\bc,-4.0);
		\filldraw[draw=black,thick,fill={rgb,1:red,0.89;green,0.1;blue,0.1}] (3.0+\code+\code*\bc,-3.0) rectangle (4.0+\code+\code*\bc,-4.0);
		% before: 5 15 8 4 9 6 10 8 3 7 13 1 12 2 11 3 
		\node[] at (0.5+\code+\code*\bc, -0.5) {$x_5$};
		\node[] at (1.5+\code+\code*\bc, -0.5) {$x_{15}$};
		\node[] at (2.5+\code+\code*\bc, -0.5) {$x_8$};
		\node[] at (3.5+\code+\code*\bc, -0.5) {$x_4$};
		\node[] at (0.5+\code+\code*\bc, -1.5) {$x_9$};
		\node[] at (1.5+\code+\code*\bc, -1.5) {$x_6$};
		\node[] at (2.5+\code+\code*\bc, -1.5) {$x_{10}$};
		\node[] at (3.5+\code+\code*\bc, -1.5) {$x_8$};
		\node[] at (0.5+\code+\code*\bc, -2.5) {$x_3$};
		\node[] at (1.5+\code+\code*\bc, -2.5) {$x_7$};
		\node[] at (2.5+\code+\code*\bc, -2.5) {$x_{13}$};
		\node[] at (3.5+\code+\code*\bc, -2.5) {$x_1$};
		\node[] at (0.5+\code+\code*\bc, -3.5) {$x_{12}$};
		\node[] at (1.5+\code+\code*\bc, -3.5) {$x_2$};
		\node[] at (2.5+\code+\code*\bc, -3.5) {$x_{11}$};
		\node[] at (3.5+\code+\code*\bc, -3.5) {$x_3$};
		\node[] at (2.0+\code+\code*\bc, -4.5) {\small{(a) before the parasitic attack}};
		% code number
		\def\code{1}
		\filldraw[draw=black,thick,fill={rgb,1:red,0.99;green,0.74;blue,0.43}] (0+\code+\code*\bc,-0) rectangle (1.0+\code+\code*\bc,-1.0);
		\filldraw[draw=black,thick,fill={rgb,1:red,0.85;green,0.85;blue,0.85}] (1.0+\code+\code*\bc,-0) rectangle (2.0+\code+\code*\bc,-1.0);
		\filldraw[draw=black,thick,fill={rgb,1:red,0.41;green,0.23;blue,0.60}] (2.0+\code+\code*\bc,-0) rectangle (3.0+\code+\code*\bc,-1.0);
		\filldraw[draw=black,thick,fill={rgb,1:red,0.69;green,0.87;blue,0.54}] (3.0+\code+\code*\bc,-0) rectangle (4.0+\code+\code*\bc,-1.0);
		\filldraw[draw=black,thick,fill={rgb,1:red,1;green,1;blue,0.60}] (0+\code+\code*\bc,-1.0) rectangle (1.0+\code+\code*\bc,-2.0);
		\filldraw[draw=black,thick,fill={rgb,1:red,1;green,0.49;blue,0}] (1.0+\code+\code*\bc,-1.0) rectangle (2.0+\code+\code*\bc,-2.0);
		\filldraw[draw=black,thick,fill={rgb,1:red,0.98;green,0.50;blue,0.44}] (2.0+\code+\code*\bc,-1.0) rectangle (3.0+\code+\code*\bc,-2.0);
		\filldraw[draw=black,thick,fill={rgb,1:red,0.41;green,0.23;blue,0.60}] (3.0+\code+\code*\bc,-1.0) rectangle (4.0+\code+\code*\bc,-2.0);
		\filldraw[draw=black,thick,fill={rgb,1:red,0.21;green,0.59;blue,0.56}] (0+\code+\code*\bc,-2.0) rectangle (1.0+\code+\code*\bc,-3.0);
		\filldraw[draw=black,thick,fill={rgb,1:red,0.79;green,0.69;blue,0.83}] (1.0+\code+\code*\bc,-2.0) rectangle (2.0+\code+\code*\bc,-3.0);
		\filldraw[draw=black,thick,fill={rgb,1:red,0.30;green,0.68;blue,0.29}] (2.0+\code+\code*\bc,-2.0) rectangle (3.0+\code+\code*\bc,-3.0);
		\filldraw[draw=black,thick,fill={rgb,1:red,0.41;green,0.80;blue,0.89}] (3.0+\code+\code*\bc,-2.0) rectangle (4.0+\code+\code*\bc,-3.0);
		\filldraw[draw=black,thick,fill={rgb,1:red,0.75;green,0.5;blue,0.17}] (0+\code+\code*\bc,-3.0) rectangle (1.0+\code+\code*\bc,-4.0);
		\filldraw[draw=black,thick,fill={rgb,1:red,0.98;green,0.60;blue,0.60}] (1.0+\code+\code*\bc,-3.0) rectangle (2.0+\code+\code*\bc,-4.0);
		\filldraw[draw=black,thick,fill={rgb,1:red,1;green,0.90;blue,0.40}] (2.0+\code+\code*\bc,-3.0) rectangle (3.0+\code+\code*\bc,-4.0);
		\filldraw[draw=black,thick,fill={rgb,1:red,0.87;green,0.76;blue,0.49}] (3.0+\code+\code*\bc,-3.0) rectangle (4.0+\code+\code*\bc,-4.0);
		% after : 5 15 8 4 9 6 10 8 29 7 13 1 21 2 11 19 
		\node[] at (0.5+\code+\code*\bc, -0.5) {$x_5$};
		\node[] at (1.5+\code+\code*\bc, -0.5) {$x_{15}$};
		\node[] at (2.5+\code+\code*\bc, -0.5) {$x_8$};
		\node[] at (3.5+\code+\code*\bc, -0.5) {$x_4$};
		\node[] at (0.5+\code+\code*\bc, -1.5) {$x_9$};
		\node[] at (1.5+\code+\code*\bc, -1.5) {$x_6$};
		\node[] at (2.5+\code+\code*\bc, -1.5) {$x_{10}$};
		\node[] at (3.5+\code+\code*\bc, -1.5) {$x_8$};
		\node[] at (0.5+\code+\code*\bc, -2.5) {$x_{29}$};
		\node[] at (1.5+\code+\code*\bc, -2.5) {$x_7$};
		\node[] at (2.5+\code+\code*\bc, -2.5) {$x_{13}$};
		\node[] at (3.5+\code+\code*\bc, -2.5) {$x_1$};
		\node[] at (0.5+\code+\code*\bc, -3.5) {$x_{21}$};
		\node[] at (1.5+\code+\code*\bc, -3.5) {$x_2$};
		\node[] at (2.5+\code+\code*\bc, -3.5) {$x_{11}$};
		\node[] at (3.5+\code+\code*\bc, -3.5) {$x_{19}$};
		\node[] at (2.0+\code+\code*\bc, -4.5) {\small{(b) after the parasitic attack}};
	\end{tikzpicture}
	\caption{\small{Partition of the environmental states induced by the code of an agent (a) before, and (b) after the parasitic attack.}}
  \label{fig:code_change}
\end{figure}

This can be appreciated in Fig. \ref{fig:code_change}, where we represent the code of an agent before and after the population's response to the
parasitic attack. This agent updated its code such that most symbols used by the parasite are avoided ($x_3$ and $x_{12}$ are changed for $x_{29}$
and $x_{21}$, respectively). On the other hand, $x_{13}$ is kept. To check whether this is an optimal solution, we manually updated the code of all
agents of the same type, changing $x_{13}$ with every other possible symbol. Indeed, using this particular symbol occupied by the parasite maximises
the population's mutual understanding. The reason for this is that, since all other symbols are occupied by more than one agent, $x_{13}$ is the one
that confuses the population the least.

Second, by drifting from the parasite's symbols, agents may update their codes in such a way that, after the update, they capture more environmental
information. This is the case of the type shown in Fig. \ref{fig:code_change}: before the update, environmental states $9$ and $16$ (see Fig.
\ref{fig:code_types}a to locate these states) were represented by $x_3$, while after the update, these states are distinguished from one another.

% original
%
% before: 4 14 7 3 8 5 9 7 2  6 12 0 11 1 10 2
% after : 4 14 7 3 8 5 9 7 28 6 12 0 20 1 10 18

Third, and most important, the information that the parasite offers can now be understood (although not entirely) by the population: the missing
information is mostly expressed using symbols that are not occupied any more by the agents of the population. This cannot be shown in the example, since
changes in the agent's codes after the response to the attack may result in an overlap with the information that the parasite captures. However, we can
manipulate the resulting configuration after the parasitic attack to show that agents now consider the information offered by the parasite. For each agent
that is not the parasite, we update its code such that $p\left(x + 16 \;\middle\vert\; \mu\right) \coloneqq p\left(x \;\middle\vert\; \mu\right)$ and then
we set $p\left(x \;\middle\vert\; \mu\right) \coloneqq 0$. In this way, we make sure that all agents capture the same aspects of $\mu$ as before the update,
without interference (all of the parasite's symbols are in the range $[1,16]$).

Now, we measure the average environmental information before and after the change. Before, the value was
$I\left(\mu \;;\; X_\Theta, X_{\Theta^\prime}\right) = 3.70$ bits, and after, $I\left(\mu \;;\; X_\Theta, X_{\Theta^\prime}\right) = 3.72$
bits. The increase is small, but we are considering one parasite only. If we introduce $8$ parasites in the population, then the increase in the
average environmental information is more significant: from $3.43$ bits to $3.73$ bits after updating the codes. It is worth noting that,
if the parasites interact between themselves, then they would try to capture not only environmental information that is not present in the
population, but also that is not captured by the other parasites.

\section{Discussion and conclusions}
\label{sec:discussion}

We have considered a scenario where a parasite is introduced in a previously evolved population, and, after convergence, we looked at the
response of the population, in one step of many in the co-evolutionary arms race. We considered one step only since, in this setting where
the agent's behaviours are not unified, the arms race will cycle continuously.

Our model shows interesting behaviour consistent with empirical observations. For instance, parasites are known to mimic the chemical
signatures utilised by the attacked host \citep{dEttorre2002, Lorenzi2014}. In this way, identification of the parasite by the population
becomes harder. We measured this property during the parasite's attack, showing that as it increased damage in the mutual understanding
of the population, it blended in. Additionally, we showed that it becomes parasitically dependent on the population, as most of the
environmental information it uses to predict the environment comes from the population.

We have also showed which properties a population may have in order to be robust against parasitic attacks. For instance, large populations
are more resilient, since its numerous members provide a solid standard from which perturbations become less significant. Another way
in which the population becomes resilient is for the population's agents to utilise synonyms. If the parasite intends to create confusion
among the population by using messages that have a different meaning for the rest of the agents, then when synonyms are present, then
they do not present any ambiguities.

% parasites preferentially attacks the most frequent type, while in this system it attacks all types, damaging the most frequent the less
%Our results showed that the parasite attacks all members of the population, although this does not always have to be the case. This is
%related to how similar the adopted conventions (by each sub-population) are. The more similar, then we expect the parasite to be able
%to reduce the whole population's mutual understanding, while if conventions are mostly dissimilar, then the parasite might be better off
%attacking a subset of the population only. The reason for this is that it would be difficult to chose conventions that confuse all agents
%altogether. In the extreme case where a parasite would choose one sub-population only in its attack, then we can speculate that this would
%be the more frequent one: small sub-populations that have a high mutual understanding would have less impact on the global mutual
%understanding.

The presence of parasites in a population can be, in the long term, a positive force \citep{Hudson2006}. For instance, they increase
the diversity of the population, which in our scenario was manifested in the creation of new types of codes by using synonyms. As we have
seen, this makes the population more robust to subsequent attacks in their co-evolution. Second, parasites are able to capture information
about the environment which is not captured by any other agent. Most of this information is not understandable by the agents until they
respond to the parasitic attack by drifting their codes.

The code drift has two effects: first, it makes the parasite easier to identify, since it is the only agent using a particular set of messages;
and second, after the messages used by the parasite are avoided, the parasite's information becomes understandable for the whole population.
Therefore, after the population recovers from the attack, agents can improve their predictions of the environment. The parasite, after the
population's response, can still perfectly predict the state of the environment, but with one major drawback: it becomes easily identifiable,
and thus the population have the possibility to take action (for instance, by avoiding interaction) when a future attack begins.

%\section{Conclusions}
%\label{sec:conclusions}
%Our model of code evolution shows complex and interesting outcomes. In this work, we make use of it to study how the introduction of parasites
%in a population can enhance the population's general functioning.
%
%Although we consider only one particular behaviour for our parasite, this study provides new insights over the beneficial effects of parasitism.
%Our results show that, in order to inflict as much damage as possible, the parasite must express information that is missing in the population
%with the most popular messages. This information would be highly valuable, providing that agents can understand it. Once the agents of the
%population drift their codes by responding the attack, they become capable of ``reading'' the parasite. Then, they can use that information
%to improve their predictions and increase their growth rate.

\footnotesize
\bibliographystyle{apalike}
\bibliography{paper}

\end{document}